\documentclass[fleqn,12pt]{wlscirep}
\usepackage[utf8]{inputenc}
\usepackage[T1]{fontenc}
\usepackage{etoolbox}

\DeclareUnicodeCharacter{2500}{?}

\usepackage{longtable}
\usepackage{comment}
\def \muJy {$\mu$Jy}

\newcommand{\apj}{Astrophys. J.}
\newcommand{\apjl}{Astrophys. J.}
\newcommand{\apjs}{Astrophys. J.}
\newcommand{\aap}{Astron. Astrophys.}
\newcommand{\mnras}{Mon. Not. R. Astron. Soc.}
\newcommand{\nat}{Nature}
\newcommand{\nata}{Nature Astronomy}
\newcommand{\araa}{Ann. Rev. Astron. Astrophys.}

\newcommand{\ssr}{Space Science Rev.}

\newcommand{\physreps}{Phys. Rep.}
\newcommand{\aaps}{Astron. Astrophy. Sup.}

\usepackage{lineno}

\usepackage{bm}
\DeclareCaptionLabelSeparator{mysep}{$\boldsymbol{|}$}
\title{\center Thermonuclear explosions on neutron stars reveal the speed of their jets}

\author{
Thomas D. Russell$^{1*}$,  
Nathalie Degenaar$^{2}$, 
Jakob van den Eijnden$^{3,4}$,
Thomas Maccarone$^{5}$,
Alexandra J. Tetarenko$^{5,6,7 \dagger}$,
Celia S\'anchez-Fern\'andez$^{8}$,
James C. A. Miller-Jones$^{9}$, \\
Erik Kuulkers$^{10}$, 
Melania Del Santo}

\affil[1]{INAF, Istituto di Astrofisica Spaziale e Fisica Cosmica, Via U. La Malfa 153, I-90146 Palermo, Italy}
\affil[2]{Anton Pannekoek Institute for Astronomy, University of Amsterdam, Science Park 904, Amsterdam 1098 XH, The Netherlands}
\affil[3]{Department of Physics, University of Warwick, Coventry CV4 7AL, UK}
\affil[4]{Department of Physics, Astrophysics, University of Oxford, Denys Wilkinson Building, Keble Road, Oxford OX1 3RG, UK}
\affil[5]{Department of Physics and Astronomy, Texas Tech University, Lubbock, TX 79409-1051, USA}

\affil[6]{Department of Physics and Astronomy, University of Lethbridge, Lethbridge, Alberta, T1K 3M4, Canada}
\affil[7]{East Asian Observatory, 660 N. A'oh\={o}k\={u} Place, University Park, Hilo, Hawaii 96720, USA}
\affil[8]{Science Operations Department, European Space Astronomy Centre (ESA/ESAC), Villanueva de la Ca\~nada, Madrid 28691, Spain}
\affil[9]{International Centre for Radio Astronomy Research, Curtin University, Bentley, WA 6102, Australia}
\affil[10]{European Space Research and Technology Centre (ESTEC), European Space Agency (ESA), Keplerlaan 1, Noordwijk 2201 AZ, The Netherlands}
\affil[$\star$]{Corresponding author. Email:thomas.russell@inaf}
\affil[$\dagger$]{Former NASA Einstein Fellow}

\begin{document}
\flushbottom
\maketitle

{\bf
\noindent
Relativistic jets are observed from accreting and cataclysmic transients throughout the Universe, and have a profound affect on their surroundings\cite{2005Natur.436..819G,2012ARA&A..50..455F}. Despite their importance, their launch mechanism is not known. For accreting neutron stars, the speed of their compact jets can reveal whether the jets are powered by magnetic fields anchored in the accretion flow \cite{1982MNRAS.199..883B} or in the star itself \cite{2016ApJ...822...33P,2022MNRAS.515.3144D}, but to-date no such measurements exist. These objects can display bright explosions on their surface due to unstable thermonuclear burning of recently accreted material, called type-I X-ray bursts \cite{2021ASSL..461..209G}, during which the mass accretion rate increases \cite{2018SSRv..214...15D,2018ApJ...867L..28F, 2020NatAs...4..541F}. Here, we report on bright flares in the jet emission for a few minutes after each X-ray burst, attributed to the increased accretion rate. With these flares, we measure the speed of a neutron star compact jet to be $v=0.38^{+0.11}_{-0.08}$c, much slower than those from black holes at similar luminosities. This discovery provides a powerful new tool in which we can determine the role that individual system properties have on the jet speed, revealing the dominant jet launching mechanism. 

}
\vspace{30pt}

The accreting neutron stars 4U~1728$-$34 and 4U~1636$-$536 (hereafter, 4U1728 and 4U1636, respectively) both exhibit frequent type-I X-ray bursts, with recurrence rates of $\approx$0.3 bursts per hour \cite{2010ApJ...724..417G,1987ApJ...319..893L}. For each of these two sources we carried out an intensive, simultaneous radio and X-ray observing campaign. The radio observations, which probe the jet emission, were taken over a three day period with the Australia Telescope Compact Array (ATCA), with $\sim$30 hours of time on source for each individual source. Radio data were recorded simultaneously at two frequency bands, 5.5 and 9\,GHz. X-ray monitoring, tracing the accretion flow, consisted of a single long observation for each source with the INTErnational Gamma-Ray Astrophysics Laboratory (INTEGRAL), lasting an entire satellite orbit of around 51 hours and ensuring strictly simultaneous monitoring with our radio campaign. During the observations, the persistent X-ray emission from both sources was consistent with being in a hard X-ray spectral state known to be associated with jets.

\renewcommand{\figurename}{Fig.}
\begin{figure}
   \centering
      { \includegraphics[width=15cm]{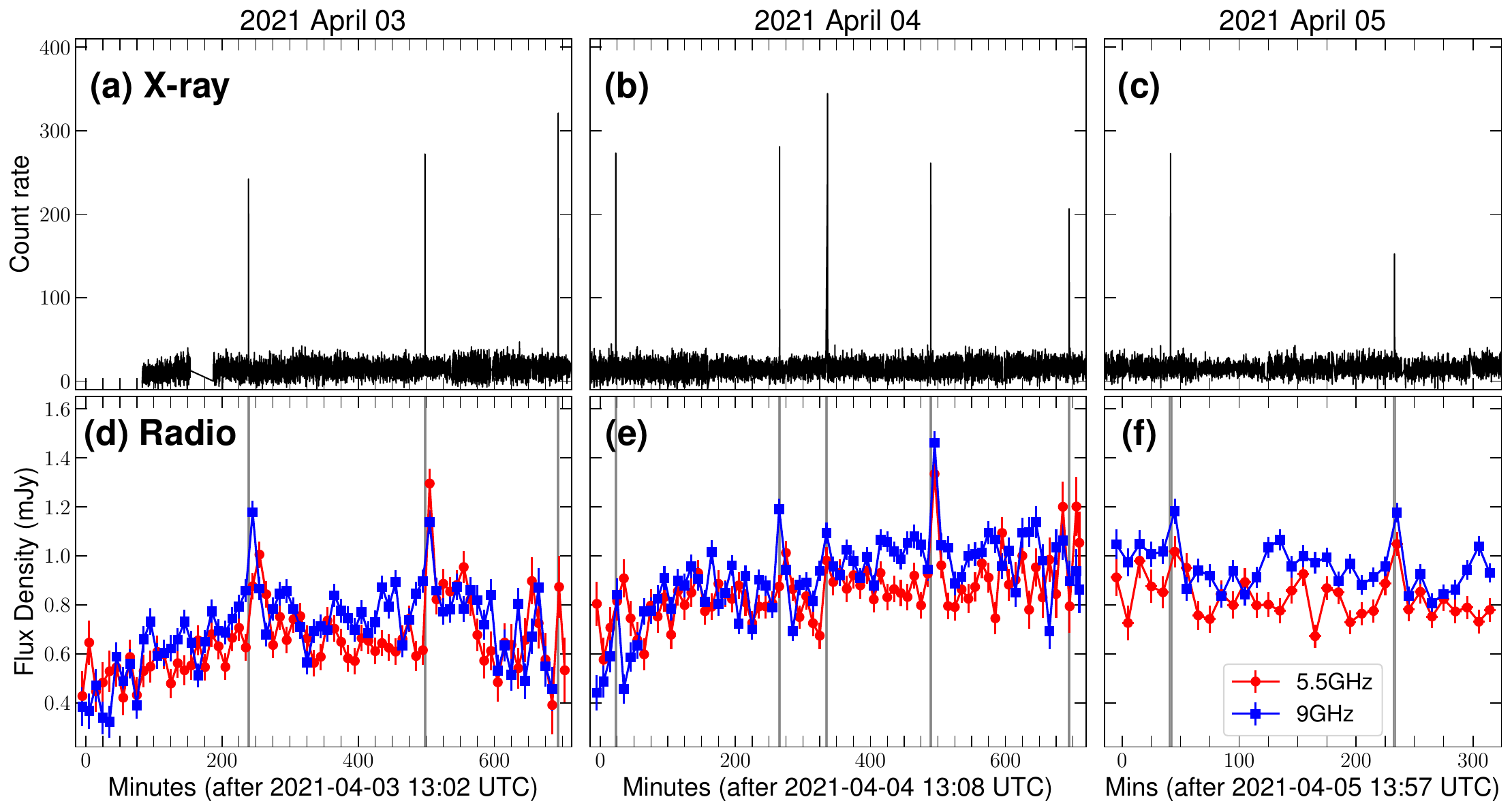}}\\

 \caption{\label{fig:lc_4u1728} Strictly simultaneous X-ray and multi-band radio light curves of 4U1728. For the X-rays, we show the 2-second 3--25\,keV count rate for each epoch  (panels a, b, c, where each panel corresponds to a different epoch). For the radio (panels d, e, f), we show the flux densities of the target during each epoch, measured at 5.5\,GHz (red circles) and 9\,GHz (blue squares) for 10-minute time-bins. Error bars show the 1-sigma uncertainties on the flux density. The timing of the X-ray bursts in the X-ray light curves are shown by the grey vertical lines in the lower panels. We detect 10 X-ray bursts coincident with our radio monitoring. For all X-ray bursts we find clearly defined radio counterparts, although for the final bursts of epoch 1 and 2 the data are not as clear due to the low source elevation and the radio observation ending close to the burst.
 }
 \end{figure}

\begin{figure}
   \centering
   
   { \includegraphics[width=12cm]{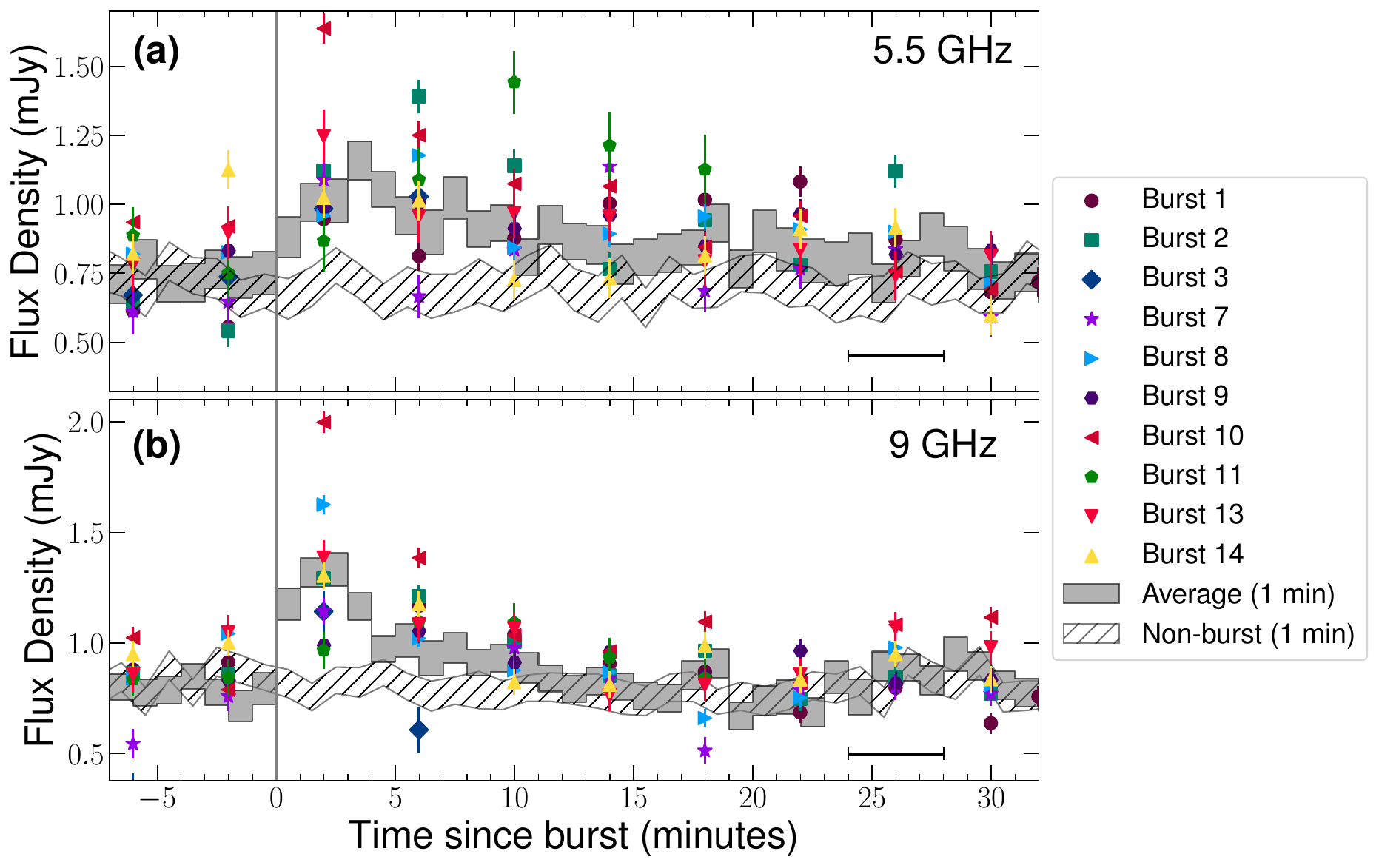}}\\

 \caption{\label{fig:bursts_4min} Radio light curves of 4U1728 showing the radio response to each of the ten individual bursts detected in the radio band. Here we show the 4-minute time intervals at 5.5\,GHz (panel a) and 9\,GHz (panel b), where the data points show the centre of the time bin. The width of each time bin is shown by the error bar in the lower right of each panel and errors on the flux density are 1-sigma. By combining all of the radio data following a burst, we show the average response at each frequency on 1-minute timescales (grey shaded area, where the height represents the uncertainties in the flux density, and the width is the duration of the time bin), as well as the average radio flux density at non-burst times (hatched area, where the width is the uncertainty), determined by randomly sampling the non-burst light curve. Bursts are numbered according to their detection in the X-ray monitoring, where 14 bursts were detected. However, only 10 bursts occurred with simultaneous radio data. Here we see that the radio response from each single burst show significant variability in brightness, and the rate at which they brighten.
 }
 \end{figure}

Fourteen type-I X-ray bursts occurred during our INTEGRAL monitoring of 4U1728, ten of which occurred when the source was visible to ATCA. In the 3--25\,keV energy band the X-ray bursts all have a similar duration of $\sim$10\,s, but differ in peak brightness by a factor $\sim$2. Following every X-ray burst exhibited by 4U1728, we detected a clear radio flare with the radio emission brightening by a factor of 1.36--1.90 in the minutes after each X-ray burst (Fig.~\ref{fig:lc_4u1728}), where the range represents the errors on the average light curves. The timing of the radio flare peak was frequency dependent; on average, the 5.5\,GHz radio emission peaked 3.5$\pm$0.5\,mins after the onset of the X-ray flare, before fading back to pre-burst levels 20--25\,mins after the X-ray burst (Fig.~\ref{fig:bursts_4min}). In contrast, at 9\,GHz, the emission peaked 2.5$\pm$0.5\,mins after the start of the burst, fading back to pre-burst levels within $12 \pm 2$\,mins. Analysis of each individual radio flare shows variability in both the rise and decay times, as well as the brightening factor (Fig.~\ref{fig:bursts_4min}). The combination of the flare arriving first at higher frequencies and strong variations between individual flares points to a jet origin of the radio enhancement. While reflection can also produce time lags at higher frequencies, reflected emission from the disk or companion star is expected to occur within a few seconds of the burst, and for the observed radio emission to arise from thermal reprocessing then the emission region required would dominate the total emission, being far brighter than even the burst itself (following arguments in \cite{2010MNRAS.404L..21C}). Furthermore, calculating the brightness temperature, $T_{\rm b}$, from the brightness and rise times of the radio flares ($T_{\rm b} > 4.5 \times 10^{9}$\,K) implies that the emission must be non-thermal in origin. The radio flare properties also suggest that the emission originates from outside of the orbit of the neutron star, and that it is unlikely to arise from a disk wind, as its speed would well exceed those typically observed from accreting neutron stars (by a factor of $\gtrsim 7$). As such, our observations imply that the radio flares associated with the type-I X-ray bursts originate from an out-flowing jet.

During their hard states, accreting neutron stars launch a partially self-absorbed compact jet, that is  characterised by a flat to inverted synchrotron spectrum that extends from the radio to the infrared band. However, at their highest mass accretion rates during the transition to their soft X-ray state the jet emission changes dramatically, where discrete, ballistic ejecta can be launched. The radio emission can show bright flares as the ejecta travel outwards and interact with their local environment \cite{2006csxs.book..381F}, with a radio spectrum that becomes steep as the peak of the self-absorbed synchrotron spectrum shifts to lower frequency as the ejecta expand. A steep radio spectrum is observed even in the case of short-lived, faint ejections from accreting neutron stars \cite{2023Natur.615...45V}. Throughout our radio monitoring, and in particular during the radio flares, the radio spectrum remained flat or inverted at all times indicating that the flares occurred within a compact jet (see Methods section).

For 4U1636, twelve bursts occurred during our INTEGRAL monitoring, with ten occurring during our radio observations. However, 4U1636 was radio faint during our observations such that we were not able to detect flaring at such low levels of emission on the short timescales required to repeat the individual burst analysis that we were able to complete for 4U1728. By combining all of the 4U1636 radio data from the onset of the X-ray burst to 20\,mins post-burst we do indeed find an enhancement of the radio emission when compared to the emission prior to and after the flares. Using the timescales of the radio flares observed from 4U1728, we place an estimate on the brightening factor of $\sim$1.2 -- 1.8 for 4U1636, similar to results from 4U1728.

Observational and theoretical evidence shows that for the duration of an X-ray burst, the mass accretion rate onto the neutron star increases \cite{2013A&A...553A..83I} by up to a factor of $\sim$10 \cite{2018ApJ...867L..28F, 2020NatAs...4..541F}. The increased influx of photons from an X-ray burst onto the accretion flow can be expected to cause two changes to the accretion flow’s properties driving the increase in the jet emission: first, hot geometrically thick, optically thin accretion flows, like those seen in hard state accretion, will be cooled via inverse Compton processes and will hence collapse geometrically, as their random thermal motions will no longer exceed their ordered orbital motions. The strong reduction of the hard X-rays in several systems during type-I bursts has been observed, supporting this picture \cite{2003MNRAS.338..189M}. Secondly, Poynting-Robertson drag effects rapidly cool plasma within the geometrically thin, optically thick disc causing the matter in the accretion flow to collapse vertically and fall in faster than the standard viscous timescale, increasing the accretion rate in the disk for the duration of the X-ray burst \cite{2005ApJ...626..364B, 2018ApJ...867L..28F, 2020NatAs...4..541F}.

In magneto-centrifugal jet launching models developed to include objects without an event horizon, such as the Blandford-Payne jet model classically assumed for weak-magnetic field neutron stars \cite{1982MNRAS.199..883B}, the magnetic field responsible for jet launching is generated by and anchored in the inner disk. In this case, the kinetic power entering a relativistic jet is typically believed to be correlated with the poloidal component of the magnetic field of the accretion disk \cite{2001ApJ...548L...9M}. The increase of the mass accretion rate from the Poynting-Robertson drag will then lead to an increase in magnetic field strength in the inner accretion disk. However, if the magnetic fields collapse inward with the gas in the disk, then the poloidal magnetic field component weakens, which would lead to a suppression of the jet power even with the increase in the accretion rate. Alternatively, if the magnetic field either remains static, or collapses more slowly than the gas in the disk, then the increase in accretion rate simply leads to an increase of kinetic power into the jet, driven by the burst, in agreement with our observations of a radio flare associated with each X-ray burst. We note that the radio brightening could also arise from magnetospheric jet mechanisms, where the neutron star's magnetic field is responsible for launching the jet \cite{2016ApJ...822...33P,2022MNRAS.515.3144D}, either in powering the jet continuously, or in the jet's brightening as it takes over from weakened magneto-centrifugal launching.

Both accreting stellar-mass black holes and neutron stars launch jets. Despite these two classes of systems having profoundly different physical characteristics, i.e., an event horizon versus a stellar surface, very few differences between the emitted jets have been identified, beyond the jets appearing to be generally more luminous in black hole systems at similar X-ray luminosities \cite{2018MNRAS.478L.132G}. In the few cases that it has been estimated, the compact jet speed from black holes has been shown to be highly relativistic, where $\beta > 0.63$ ($\beta = v/{\rm c}$, where $v$ is velocity and c is the speed of light\cite{2010MNRAS.404L..21C,2019ApJ...887...21S,tetarenkoa19,2021MNRAS.504.3862T,2022ApJ...925..189Z}), but the most robust constraints generally exceed $\beta = 0.9$. Discrete, ballistic ejecta can also be launched from  black holes as they approach high mass accretion rates ($\sim$a few to tens of percent of the Eddington luminosity), as well as highly-accreting, persistent (Z-type) neutron star systems. The speeds of these discrete ejecta can be variable, in the few cases they have been detected in neutron star systems, the ejecta travel outwards at mildly relativistic speeds \cite{2001ApJ...558..283F,2013MNRAS.435L..48S} but for black hole systems, the ejecta can be either highly- or mildly-relativistic \cite{1994Natur.371...46M,2021MNRAS.505.3393W}. However, to date, no robust measurements have been made for the compact jets from the transient (atoll-type) neutron star systems, for comparison with the fast compact jets from black holes. While it is often assumed that the compact jets from accreting neutron stars should be slower than their black hole counterparts due to a lower escape velocity, the lack of any observational evidence to support this has prevented a clear comparison of jet launching between these objects. The detection of radio bursts allows for such a comparison for the very first time; different radio frequencies probe emission from a narrow range of distances downstream in the jet, where higher frequencies arise from the more compact regions of the jet, closer to the compact object \cite{1979ApJ...232...34B}. As such, by measuring the time delay between the X-ray burst and the radio flare at each radio frequency (i.e., the time lags), we are able to measure the compact jet speed\cite{tetarenkoa19} from 4U1728 (see Methods section for more details). We measure a jet speed of $v =0.38^{+0.11}_{-0.08}$\,c. This is the first time robust constraints have been placed on the compact jet speed from a confirmed accreting neutron star. This result is much slower than measurements made for the compact jets from black hole systems, and is similar to the expected escape velocity for a neutron star.

The discovery of jet flares associated with type-I X-ray bursts presents the first indication that the disk magnetic field collapses more slowly than the gas in the disk and the first measurement of the compact jet speed from a neutron star system. Neutron stars offer a unique laboratory to study jet launching; the spin and magnetic field can often be measured in these systems, both of which are believed to fundamentally affect jet launching \cite{1977MNRAS.179..433B, 2016ApJ...822...33P}. As such, the discovery of jet flares associated with type-I X-ray bursts offers a completely new and robust way in which we can tie the jet speeds to fundamental jet powering properties. This phenomenon opens a new window into the dominant jet launching mechanism; measuring the jet speeds from a large enough population with a variety of mass and spin rates will reveal if there is a direct connection to these properties, and how they compare to black holes. For example, magnetospheric jet models predict a connection between the compact jet speed and neutron star spin \cite{2016ApJ...822...33P,2022MNRAS.515.3144D}, while testing how the jet speed is affected by the neutron star mass can reveal if the jets are connected to the escape velocity of the system \cite{1999PhR...311..225L}. In addition, a more detailed study of how individual X-ray bursts, which produce different increases to the mass accretion rate, impact the jets will significantly advance our knowledge of jet launching. In particular, the jet response at higher frequencies would enable even tighter constraints on the jet speed. Therefore, upgrades to existing facilities such as ATCA, ALMA, SMA, NOEMA, and new radio and millimetre telescopes such as the Next-Generation Very Large Array, the Africa Millimetre Telescope, and the Square Kilometre Array will provide the sensitivity and frequency coverage required to analyse single radio bursts in much greater detail from many systems.

\newpage

\newpage
\section{Methods}
\label{sec:methods}

\subsection{X-ray observations}
Both 4U1728 and 4U1636 alternate between hard and soft X-ray states on a time scale of weeks \cite{2014MNRAS.443.3270M}. To ensure that a compact radio jet was present, our ATCA and INTEGRAL observations were aimed to be performed in the hard X-ray spectral state. To this end, we triggered our campaigns based on the spectral state evolution seen from the publicly available X-ray monitoring light curves of the Monitor of All-sky X-ray Image mission (MAXI, 2--20 keV\cite{2009PASJ...61..999M}) and the Burst Alert Telescope (BAT, 15--50 keV\cite{2013ApJS..209...14K}) on board the Swift satellite \cite{2004ApJ...611.1005G}.

\subsubsection{INTEGRAL data reduction and analysis}
\label{subsect:inte}

INTEGRAL \cite{2003A&A...411L...1W,2021NewAR..9301629K} observed 4U1728 from 2021-04-03 14:26 UT to 2021-04-05 23:26 UT (MJD~59,307.6014 -- MJD~59,309.9764; obsID 18700020001), and 4U1636 from 2021-10-04 04:10 UT to 2021-10-05 12:05 UT (MJD~59,491.1736 -- MJD~59,492.5035; obsID 18700050001), for total exposure times of 186.6 and 182.6 kiloseconds (corresponding to 51.83 and 50.72 hours, respectively). These observations took place in satellite revolutions 2350 (4U1728) and 2419 (4U1636). The observations were scheduled in the Hexagonal dither pattern (HEX), consisting of one source-on-axis pointing, six source-off-axis pointings 2.17$^\circ$ apart in a hexagonal pattern centred on the nominal target position. Each target was monitored almost continuously for an entire revolution, except for the small slews from pointing to pointing in the pattern, which resulted in periodic 2-minute gaps in the exposures. 

In this work we analyse the data provided by INTEGRAL's instruments JEM-X \cite{2003A&A...411L.231L} and IBIS \cite{2003A&A...411L.131U}. JEM-X, the Joint European X-ray Monitor for X-rays, is sensitive in the 3--35 keV energy range, with a timing resolution capable of reaching 122\,$\mu$s and an angular resolution of $3^{\prime}$. It consists of two identical units X1 and X2, which operated simultaneously.  ISGRI, the INTEGRAL Soft Gamma-Ray detector of the IBIS (Imager on-Board the INTEGRAL Satellite) instrument is sensitive in the 15\,keV to 1\,MeV range, and provides an angular resolution of $12^{\prime}$. The data were reduced using the INTEGRAL Offline Science Analysis (OSA) v.11.2 \cite{2003A&A...411L..53C}, provided by the ISDC Data Centre for Astrophysics (\url{http://www.isdc.unige.ch/}, using standard reduction steps. \\

\subsubsection{X-ray burst detection}
X-ray bursts from 4U1728 and 4U1636 were identified on the 1-s binned JEM-X light curves extracted in the 3--25~keV band at the science window observation level (3393 and 3320 seconds duration, respectively). X-ray light curves are shown in the top panel of Extended Data Figure~\ref{fig:bursttimes}. We computed the mean and standard deviation of the source count rate during each science window, and recursively compared the source count rate with these values for each time bin. When a bin rate exceeded 6$\sigma$ of the persistent level emission we identified the potential onset of an X-ray burst. The peak of the burst and subsequent burst decay, which we fitted with an exponential function, were identified in the system light curve to verify that the rise and decay profile were consistent with those of a type-I X-ray burst, and to discard potential disturbances in the system light curve. The burst detection was further confirmed by extracting the image of the field within a time interval restricted to the burst duration, to discard contamination to the source light curve by an event originating in another source in the field of view (FOV). The burst search resulted in a sample of 14 X-ray bursts for 4U1728 and 12 X-ray bursts for 4U1636.

\subsubsection{X-ray spectral analysis}
 \textbf{Time-resolved analysis of bursts:} To derive the main properties of each burst, such as the temperature, emitting radius and total radiated energy, we extracted JEM-X and IBIS/ISGRI time-resolved spectra, varying exposure times as a function of the count rate to ensure similar total numbers of counts in each spectrum. On the time scales relevant for our current analysis, the source was not detected by IBIS/ISGRI. For the time-resolved analysis we therefore only fitted the JEM-X spectra of individual bursts.

We fitted the burst emission with an absorbed black body model (\textsc{bbodyrad}) after accounting for the underlying persistent accretion emission. To this end, we extracted JEM-X and ISGRI time-averaged spectra for both 4U1728 and 4U1636 from all data that did not contain a burst. To build these spectra, we excluded an interval of [-20, 200] seconds around the X-ray burst peak of 4U1728 and  [-20, 250] seconds around the (generally longer) 4U1636 bursts.
We further excluded the first six science windows of each revolution to minimise the possibility of noise in the source spectrum due to instrument activation after perigee passage. After making these selections, the total exposure time of the persistent emission spectra was 130.5~ks for 4U1728 and 126.9~ks for 4U1636.

The persistent emission parameters during the time-resolved analysis were fixed to the values obtained from fitting the pre-burst spectra (see below), because the data quality and restricted energy range (only JEM-X) did not allow the persistent emission parameters to be variable during the bursts \cite{2013ApJ...772...94W}.  

~\\
\noindent
\textbf{Fits to persistent emission spectra:}
To characterise the persistent accretion emission, we performed joint fits to JEM-X (5–-20\,keV)  and ISGRI (40--100\,keV), where the instruments are well calibrated. Spectral data of each source were fit using \textsc{XSpec} version 12.11.0 \cite{xspec}. A multiplicative constant factor was included in the fits to take into account the uncertainty in the cross-calibration of the instruments. A systematic error of 2\% was applied to the JEM-X/ISGRI spectra which corresponds to the current uncertainty in the response matrix. Uncertainties in the results are provided at a 90\% confidence level for a single parameter. 

The average spectrum of 4U1728 was fitted with a model consisting of a thermal Comptonisation component and a hard power-law tail,  modified by the interstellar absorption in the direction of the source ($N_H=2.6\times10^{22}$ cm$^{-2}$)\cite{2013ApJ...772...94W}, for which we used the \textsc{tbabs} model \cite{2000ApJ...542..914W}, available in \textsc{XSpec}, adopting \textsc{wilms} abundances \cite{2000ApJ...542..914W} and \textsc{vern} \cite{1995A&AS..109..125V} cross sections. The spectrum does not display residuals at soft X-ray energies that would be indicative of thermal emission from the inner regions of the accretion disk and/or neutron star vicinity. The thermal Comptonisation component was taken into account using a phenomenological cutoff power-law model, \textsc{cutoffpl} ($\chi^2 / {\rm d.o.f.}$ = 20.56/22; shown in Extended Data Table~\ref{tab:pers_cutoff}), and the more physical \textsc{nthcomp} model ($\chi^2 / {\rm d.o.f.}$ = 22.34/22; shown in Extended Data Table~\ref{tab:pers_nthc}). In both cases we described the hard power law tail using the \textsc{powerlaw} model. We note that when fitting the Comptonised emission using the \textsc{nthcomp} model, the normalization of the hard power-law tail is poorly constrained. Hence, although a priori the $\chi^2 / {\rm d.o.f.}$ statistics could favour the more physical  \textsc{nthcomp+powerlaw} model, we have used the  \textsc{cutoffpl+powerlaw} model to describe the persistent emission in our fits to the burst spectra. 

The average spectrum of 4U1636 was consistent with a thermal Comptonisation model, modified by the interstellar absorption in the direction of the source ($N_H=4.4\times10^{21}$ cm$^{-2}$)\cite{2022ApJ...935..154G}. The spectrum does not display residuals that indicate the need to include a component at soft X-ray energies. The thermal Comptonisation component was fit with the cutoff powerlaw, \textsc{cutoffpl} ($\chi^2 / {\rm d.o.f.}$ = 12.86/11; Extended Data Table~\ref{tab:pers_cutoff}), and \textsc{nthcomp} ($\chi^2 / {\rm d.o.f.}$ = 8.94/11; Extended Data Table~\ref{tab:pers_nthc}) models available in \textsc{XSpec}. As in the case for 4U1728 we chose the \textsc{cutoffpl} model to fit the data.  We observe some residuals above 70 keV which we tried to model by adding an additional powerlaw component \textsc{powerlaw}. This resulted in worse statistics but gave similar results to those obtained for the Comptonisation-only model ($\chi^2 / {\rm d.o.f.}$ = 12.2/9) so the \textsc{powerlaw} was not included in our further analysis.

~\\
\noindent
\textbf{Fits to the burst spectra:} The results of our time-resolved spectroscopy of the bursts, carried out as described above, are listed in Extended Data Table~\ref{tab:resolved_1728} for 4U1728 and in the Supplementary Information for 4U1636. For both sources, the temperatures near the burst peak are found to be $\sim$2.5--3\,keV, and decrease to $\sim$1--1.5\,keV as the burst proceeds. We measure burst fluences of $\sim 1 - 3 \times 10^{-6}$\,erg\,cm$^{-2}$ for both sources, which implies total radiated burst energies of $\sim10^{39}$\,erg. Both the observed cooling and total radiated energy output are characteristic of (normal) type-I X-ray bursts \cite{gal08}.

Time-resolved spectroscopy of the INTEGRAL data does not reveal if the bursts exhibited photo-spheric radius expansion (PRE). Based on the duration of the bursts we observed ($\sim$10\,s for 4U1728 versus $\sim$1~min for 4U1636), we suspect that the 4U1728 bursts result from the ignition of a pure He layer, which are bursts that often exhibit PRE. The longer bursts of 4U1636 are more likely caused by ignition in a mixed He/H layer, which often do not display PRE \cite{gal08}.
Studying hundreds of bursts observed with RXTE, it was found that the average increase in the persistent accretion emission (hence presumably the mass-accretion rate) was similar for non-PRE bursts and PRE bursts, but that for the latter higher values could be recorded for individual bursts \cite{2015ApJ...801...60W}.

\subsection{Radio observations}

 \subsubsection{4U 1728$-$34}
We observed 4U1728 with ATCA over three epochs in early June 2021, under project code C3433. Epoch 1 was carried out between 2021-04-03 12:44:49 UT and 2021-04-04 01:05:59 UT (MJD 59,307.5311 -- MJD 59,308.0458), epoch 2 ran from 2021-04-04 13:17:40 UT to 2021-04-05 01:04:29 UT (MJD 59,308.5539 -- 59,309.0448), while the shorter epoch 3 was taken between 2021-04-05 13:08:59 UT and 2021-04-05 18:58:39 UT (MJD 59,309.5479 -- 59,309.7907). For all observations, ATCA was in its extended 6\,km (6D) configuration (\url{https://www.narrabri.atnf.csiro.au/operations/array_configurations/configurations.html}). Data were recorded simultaneously in two sub-bands with central frequencies of 5.5 and 9\,GHz, with 2\,GHz of bandwidth at each central frequency. Each sub-band was composed of 2048 1-MHz channels. We used PKS~1934$-$638 for bandpass and flux density calibration, and the nearby (3$^\circ$ away) source B1714$-$336 was used for phase calibration. Using the Common Astronomy Software Applications (\textsc{casa} version 5.1.2\cite{2022PASP..134k4501C}), data were first edited for radio frequency interference (RFI) and systematic issues, before being calibrated and imaged following standard procedures
(see \url{https://casaguides.nrao.edu/}).

\subsubsection{4U 1636$-$536}
4U1636 was also observed by ATCA over three epochs in October 2021 under project code C3433. Observations occurred between 2021-10-04 00:53:29 UT and 2021-10-04 13:30:09 UT (epoch 1; MJD~59,491.03714120 -- MJD~59,491.56260417), 2021-10-04 22:26:20 UT and 2021-10-05 13:17:00 UT (epoch 2; MJD~59,491.93495370 -- MJD~59,492.55347222), and 2021-10-05 22:13:19 UT and 2021-10-06 06:02:29 UT (epoch 3; MJD~59,492.92591435 --MJD~59,493.25172454). ATCA was in its extended 6\,km (6A) configuration throughout all observations. Data were taken at central frequencies of 5.5 and 9\,GHz, with 2\,GHz of bandwidth at each frequency, comprised of 2048 1-MHz channels. PKS~1934$-$638 was used for bandpass and flux density calibration, while B1657$-$56 (4.18$^\circ$ away) was used for phase calibration. Data were edited and calibrated following standard procedures in \textsc{casa}.

\subsubsection{Imaging and variability analysis}
Data were initially imaged with the \texttt{CLEAN} task in \textsc{casa}. Initial imaging was carried out using a Briggs robust parameter of 0 to balance sensitivity and resolution, while mitigating effects from any side-lobes present in the images. 4U1728 had an average flux density of $788 \pm 6$\,\muJy\ at 5\,GHz and $880 \pm 6$\,\muJy\ at 9\,GHz, corresponding to a radio spectral index, $\alpha$, of $0.23 \pm 0.06$. The measured flat/slightly inverted radio spectrum is consistent with emission from a self-absorbed compact jet \cite{2006csxs.book..381F}. 4U1636 was much more radio faint, where we measured an average flux density of $45 \pm 4$\,\muJy\ and $55 \pm 8$\,\muJy\ at 5.5 and 9\,GHz, respectively, where $\alpha = 0.4 \pm 0.5$, also consistent with emission from a compact jet. 
 
To analyse the short-timescale variability of the sources, we first removed the inner 10kilolambda baselines to remove any diffuse emission in the field. Second, we subtracted all sources in the field (in the $uv$-plane) using the \textsc{casa} task \texttt{uvsub}, leaving only the target, a nearby check source to ensure that the radio flares were intrinsic to the source, and a bright source towards the edge of our images. The bright source was left in as we were not able to extract it out sufficiently such that on short timescales significant noise and variability was introduced into the target and check source light curves as the bright source moved through the sidelobes in the telescope response. Keeping the bright source in the $uv$-data well accounted for such effects (as shown by the stability of the check source in Extended Data Figure~\ref{fig:lc_4u1728}). For 4U1728, the additional two sources were located at positions of right ascension (R.A.) of $17^{\rm h}31^{\rm m}48.5^{\rm s}$ and declination (Dec.) of $-33^{\circ}46^{\prime}56.1^{\prime \prime}$ (check source) and R.A. $17^{\rm h}31^{\rm m}15.0^{\rm s}$ and Dec. $-33^{\circ}51^{\prime}16.3^{\prime \prime}$ (bright source).  For 4U1636, we also extracted all sources in the field except a check source (located at R.A. $16^{\rm h}41^{\rm m}12.9^{\rm s}$ and Dec. $-53^{\circ}46^{\prime}00.7^{\prime \prime}$) and a bright source (located at $16^{\rm h}40^{\rm m}38.5^{\rm s}$ and Dec. $-53^{\circ}45^{\prime}35.05^{\prime \prime}$).

Earlier work on 4U1728 has presented radio observations taken at the time of type-I X-ray bursts \cite{2003MNRAS.342L..67M}. That work was carried out prior to the upgrade to the ATCA radio telescope, meaning that the observations were sensitivity limited, especially at short timescales. As such, \cite{2003MNRAS.342L..67M} were not able to detect any effects on the radio emission, but their limits are larger than the effect detected in our study.

For 4U1636, the radio source was a factor of $\approx$17 fainter than 4U1728, such that we were unable to significantly detect any radio variability on short timescales for single bursts (due to sensitivity limitations). As such, we searched for a radio response to the bursts by stacking the burst and non-burst times together, where the burst duration was taken from the 4U1728 results. Stacking the two (5.5 and 9\,GHz) bands provides average flux densities of $68 \pm 9$\,\muJy\ in the 20\,mins after each X-ray burst, and $46 \pm 4$\,\muJy\ outside of the burst times. Although we were not able to measure such a clear radio brightening for each individual X-ray burst from 4U1636, our results do indicate that there was indeed radio flaring associated with the X-ray bursts. Despite the limitations with sensitivity, we place an estimate on the brightening factor of $\sim$1.2 -- 1.8 for this source, similar to results from 4U1728.

Radio emission can arise from either thermal or non-thermal processes, which could originate from either a jet, the accretion flow, or wind emission. However, from our radio light curves, we measure a lower-limit on the brightness temperature of $T_{\rm b} > 4.5 \times 10^{9}$\,K (assuming a distance to source, D, of $5.2 \pm 0.5$\,kpc\cite{gal08}), indicating that the emission must arise from non-thermal processes \cite{1979rpa..book.....R,1992hea..book.....L}. In addition, assuming that the inverse Compton limit is not violated, the rise and brightness of the radio flares suggests a minimum size scale of the emission that is beyond the orbital period expected for 4U1728 \cite{2010ApJ...724..417G,2013ApJ...768..184H,2019A&A...632A..40D,2020MNRAS.495L..37V,2023MNRAS.525.2509V}. Taking the minimum size scale limit, we also find that for the emission to originate from a disk wind, the wind must be travelling at a speed well in excess of those typically observed from accreting neutron stars, by a factor of at least seven \cite{2016AN....337..368D}. As such, the timescales and brightness of the radio flares indicate a jet origin of the emission. Throughout our observations, the radio spectrum remained flat to inverted, and did not show a steepening of the spectrum at any time (Extended Data Figure~\ref{fig:radiostack}). This evolution is typical of compact jet emission, and suggests that these flares do not arise from ballistic ejection events.

Assuming equipartition between the magnetic field and the electrons, and one proton per electron, from the observed (average) brightening and flare times we estimate a minimum energy of $\sim (2-3) \times 10^{37}$\,erg required to produce the jet flares (Following equations 9.2, 9.3, and 9.4 presented in \cite{2006csxs.book..381F}). This corresponds to a mean jet power of $\sim 1\times 10^{35}$\,erg\,s$^{-1}$ with an equipartition magnetic field of $\sim$1\,G. If we assume that the flares are indeed optically-thick, we also find similar estimates (see Section 3 in \cite{2019MNRAS.489.4836F}).

\subsubsection{Measuring and modelling multi-band radio time-lags in 4U1728}
\label{sec:timelags}
To search for time-lags between the burst signals in different electromagnetic frequency bands from 4U1728, we first identify segments of the light curves for which each burst is contributing enhanced emission over the quiescent level in both the radio and X-ray light curves (see shaded regions in Extended Data Figure~\ref{fig:bursttimes}, where each burst is labelled according to Extended Data Table~\ref{tab:resolved_1728}). As we wish to remain consistent across all bursts, we define the burst segments as beginning right before the respective X-ray burst until +120 minutes after the X-ray burst (with the exception of Burst 8 (B8), where we truncate the burst segment to +60 minutes after the X-ray burst in order to avoid overlap with the next X-ray burst).
For each burst segment, we then created cross-correlation functions with these respective time segments as inputs, comparing both the X-ray to radio and 9\,GHz to 5.5\,GHz signals (using the z-transformed discrete correlation function algorithm [ZDCF]\cite{alex97,alex13a}). Further, we also repeated this procedure for an average 'stacked' version of all of the 9\,GHz to 5.5\,GHz bursts (see Extended Data Figure~\ref{fig:radiostack} and \ref{fig:lag_zdcf}). The resulting measured time-lags are shown in Extended Data Figure~\ref{fig:lags} and Extended Data Table~\ref{tab:stacked_lags}. Note that we do not consider Burst 3 and Burst 11 in our cross-correlation analysis of single bursts, as they occur close to the end of the radio observing window on that date and the radio signal is cut off before the flare decays. However, they are included in our stacked analysis (Fig.~\ref{fig:bursts_4min}).

As our measured time-lags have an electromagnetic frequency dependence, which is characteristic of out-flowing jet emission, we choose to implement the model used in \cite{tetarenkoa19}, appropriate for modelling the time-lags from a compact jet.
We only use the more statistically robust lags from the stacked signals of all the bursts for our modelling analysis (see Extended Data Table~\ref{tab:stacked_lags}; although the individual burst lags are consistent within uncertainties of the stacked values). In this model, we assume a continuous, conical, constant bulk Lorentz factor jet, where the X-ray emission originates in a region close to the neutron star and information from this region propagates down the jet axis towards the radio emission regions. Therefore, the time-lags can be represented as:

\begin{equation}
\tau_{\rm lag}=
\frac{z_{{\rm norm}}}{\beta\,c}\,(\xi_2-\xi_1)\,(1-\beta\cos\theta)\,.
\label{eq:lag}
\end{equation}

Here, $\beta$ represents the bulk jet speed (units of $v/c$, where $c$ indicates the speed of light, and the bulk Lorentz factor, $\Gamma=(1-\beta^2)^{-1/2}$) and $\theta$ represents the inclination angle of the jet axis to our line of sight. To relate the jet size scale to the electromagnetic frequency scale, we use the metric $z=z_{\rm norm}\xi=z_{\rm norm}{\nu}^{-1}$ ($\xi_{\rm 1}$ and $\xi_{\rm 2}$ are the inverse of the electromagnetic frequencies between which the time-lags were measured). Note that we introduce the $\xi$ parameter in the model formalism implemented in this work, rather than directly use the electromagnetic frequencies, as was done in \cite{tetarenkoa19}. We do this as we wish to simultaneously model X-ray to radio lags (at both radio bands), as well as lags between the 9\,GHz and 5.5\,GHz radio bands, and the $\xi$ formalism allows us to place all of these lags on a common axis of $\xi_2-\xi_1$). We also note that the $z \propto \nu^{-1}$ follows the standard jet model of \cite{1979ApJ...232...34B}, well-fitting the data. To obtain a more physically motivated prior for the jet size scale normalization parameter in our modelling, we use a one-zone synchrotron model \cite{chat11,ryblig} to express this normalization in terms of observationally measurable parameters of the system. Following the procedure outlined in Appendix C of \cite{rusellt20}, we can relate this normalization to an estimate of the size-scale of the emitting region at the jet base, which is defined primarily by the shape of the synchrotron spectrum.

\begin{equation}
    z_{\rm norm}=z_{\rm base} \nu_{\rm base} = C_1 S_{\mathrm{\nu,b}}^{(p+6)/(2p+13)} \nu_{\mathrm{b}}^{-1} D^{\frac{2p+12}{2p+13}} \nu_{\rm base}
    \label{eq:rf}
\end{equation}

Here, $\nu_{\mathrm{b}}$ is the synchrotron break frequency in the broad-band spectrum, $S_{\mathrm{\nu,b}}$ is the flux density at the break, $D$ is the distance to the source, $p$ is the power law index of the electron energy distribution (where $p$ is related to the optically thin synchrotron spectral index via $p=-2 \alpha_{\rm thin}+1$), $C_1$ is a constant term dependent on properties of the non-thermal electron population and fundamental constants, where we assume that the energy of the non-thermal electrons equals the magnetic energy density, and average over an isotropic distribution of pitch angles for the electrons, and assume $\nu_{\rm base}=\nu_{\mathrm{b}}$. To cross-check this jet size-scale method, we compare our estimate to the benchmark value presented in \cite{2020MNRAS.499..957M}. In that work, the neutron star jet base size-scale was estimated to be $\sim6\times10^8\,{\rm cm}$, based on scaling the known jet base size-scale in stellar-mass black holes (calculated from the measured $\sim100$\,ms time-lag between X-ray and optical/infrared emission found in several stellar-mass black hole systems) down by a factor of 5 due to the typical compact object mass difference. Our modelling yields a neutron star jet base size-scale of $\sim1.3\times10^8\,{\rm cm}$, which is reasonably consistent with the estimate from \cite{2020MNRAS.499..957M}, validating the jet size-scale used in our modelling. We note that any possible signal from the counter-jet, on top of the brighter approaching jet, will not affect our lag measurements as the emissions from the approaching and counter-jet have been shown to arrive at the observer at the same time, assuming both jets are symmetric \cite{2022MNRAS.517L..76M}.

We fit Equation~\ref{eq:lag} to our measured time-lags using a Markov Chain Monte-Carlo algorithm (MCMC \cite{for2013}). In our fitting runs, we allow the bulk Lorentz factor ($\Gamma$) to be a free parameter, while we sample from known distributions of inclination angle (uniform distribution in $\cos(\theta)$, where $\theta=25-53^\circ$ \cite{wang19}), synchrotron break frequency (normal distribution, $\nu_{\mathrm{b}}=(6\pm5)\times10^{13}$ Hz\cite{diaztrigo17}), flux density at the break (normal distribution, $S_{\mathrm{\nu,b}}=4\pm2$ mJy\cite{diaztrigo17}), distance (normal distribution, $D=5.2\pm0.5$\,kpc\cite{gal08}), and optically thin synchrotron spectral index (normal distribution, $\alpha_{\rm thin}=-0.8\pm0.7$\cite{diaztrigo17}). We note that for definite positive priors, such as the flux density and synchrotron break frequency, negative values were prohibited. The best-fit result is taken as the median of the one-dimensional posterior distributions and the uncertainties are reported as the range between the median and the 15th percentile (-), and the 85th percentile and the median (+), corresponding approximately to $1\sigma$ errors. Our best-fit bulk Lorentz factor  is $\Gamma=1.08^{+0.06}_{-0.03}$, corresponding to a jet velocity of $\beta = 0.38^{+0.11}_{-0.08} \, v/{\rm c}$, and the best-fit model is overlaid on the measured lags in Extended Data Figure~\ref{fig:lagfit}. We note that we choose to fit for the bulk Lorentz factor $\Gamma$, rather than the dimensionless jet speed $\beta$ (where $\Gamma = (1-\beta^2)^{-1/2}$), as it is computationally easier to avoid the hard limits on $\beta$. Using these best-fit values, the ratio of the approaching to counter-jet flux density is dependent upon jet speed ($\beta$), inclination angle ($\theta$), and spectral index ($\alpha$) according to, $\frac{F_{\rm app}}{F_{\rm counter}}=\left(\frac{1+\beta\cos\theta}{1-\beta\cos\theta}\right)^{2-\alpha}$. Inputting our best-fit values for 4U1728, with a flat spectrum ($\alpha=0$) yields $\frac{F_{\rm app}}{F_{\rm counter}}\sim2-4$.

\makeatletter
\apptocmd{\thebibliography}{\global\c@NAT@ctr 30\relax}{}{}
\makeatother

\subsection*{Acknowledgments}
We thank the anonymous referees for their excellent comments and suggestions on the manuscript. This research was made possible by an XS grant OCENW.XS3.096 from the Netherlands Organisation for Scientific Research (NWO), awarded to ND. We thank participants of the ISSI Beijing, Charleston, and Vasto meetings for very useful discussions. This research was partly based on observations with INTEGRAL, an ESA project with instruments and science data centre funded by ESA member states (especially the PI countries: Denmark, France, Germany, Italy, Switzerland, Spain) and with the participation of Russia and the USA. We thank the INTEGRAL staff for coordinating the X-ray observations possible. We also thank Jamie Stevens and ATCA staff for scheduling the ATCA observations. ATCA is part of the Australia Telescope National Facility (\url{https://ror.org/05qajvd42}) which is funded by the Australian Government for operation as a National Facility managed by CSIRO. We acknowledge the Gomeroi people as the Traditional Owners of the ATCA observatory site. TDR acknowledges support as an INAF IAF research fellow. AJT acknowledges partial support for this work was provided by NASA through the NASA Hubble Fellowship grant \#HST--HF2--51494.001 awarded by the Space Telescope Science Institute, which is operated by the Association of Universities for Research in Astronomy, Inc., for NASA, under contract NAS5--26555. 

\subsection*{Author contributions statement} TDR, ND, JvdE, JCAM-J, and TM coordinated the radio and X-ray observations. CSF analysed the X-ray data. TDR carried out and analysed the radio observations, with help from JCAM-J. AJT performed the time lag analysis. TDR, ND, JvdE, TM, AJT, and CSF interpreted the results. All authors made significant contributions to the writing of the manuscript. ND and JvdE wrote the NWO XS grant to secure ATCA observing time. TM and ND worked on the original idea of looking for a jet response to thermonuclear bursts.

\subsection*{Competing Interest Statement}

The authors declare no competing interests

\subsection*{Data availability}
 Data from INTEGRAL are publicly available online (\url{https://www.isdc.unige.ch/integral/archive}). Raw ATCA data are provided at the Australia Telescope Online Archive (\url{https://atoa.atnf.csiro.au/query.jsp}), under project code C3433. Calibrated radio and X-ray light curve data (as shown in Figure 1 and 2, and Extended Data Figures 1 and 2) are available online at \url{https://github.com/russell1/jet-burst-data.git}. Radio timing scripts are provided online at \url{https://github.com/tetarenk/AstroCompute_Scripts}. Calibrated measurement sets as well as time-lag analysis scripts (Extended Data Figures 3, 4, and 5) can be requested from the corresponding author.

\subsection*{Corresponding author}

All correspondence, including requests for data and scripts, and reprints and permissions should be sent to T. D. Russell (email: thomas.russell@inaf.it).

\clearpage
\newpage

\renewcommand*\familydefault{\sfdefault} 

\renewcommand{\figurename}{Extended Data Figure}
\renewcommand{\tablename}{Extended Data Table}
\setcounter{table}{0}
\setcounter{figure}{0}

\begin{table}
\begin{center}
\caption{Results from fitting the persistent X-ray spectrum to a phenomenological model (using \textsc{cutoffpl+powerlaw}). For both the \textsc{cutoffpl} (cut) and \textsc{powerlaw} (PL) model, the flux, $F$, is in units of erg\,cm$^{-2}$\,s$^{-1}$, $\Gamma_{\rm X}$ is the photon index of the powerlaw, and $K$ is the normalisation in photons\,keV$^{-1}$\,cm$^{-2}$\,s$^{-1}$ at 1\,keV. $E$ is the energy of the \textsc{cutoffpl} in units of keV.}
\begin{tabular}{ l c c }
Model: & \multicolumn{2}{c}{\textsc{cutoffpl+powerlaw}} \\
\hline
Parameter      & 4U1728 & 4U1636 \\
\hline
$\Gamma_{\rm cX,ut}$      & $1.8_{-0.1}^{+0.1}$ & $1.83_{-0.09}^{+0.09}$ \\
$E_{\rm cut}$           & $11_{-1}^{+1}$    & $96_{-14}^{+20}$   \\
$K_{\rm cut}$        & $1.2_{-0.2}^{+0.2}$  & $0.30_{-0.01}^{+0.01}$    \\
$F_{\rm cut}(\times 10^{-9})$         & $3.89_{-0.1}^{+0.03}$ & $2.1_{-0.1}^{+0.1}$\\
\hline
$\Gamma_{\rm X,PL}$       & $1.7_{-0.1}^{+0.1}$ & --\\
$K_{\rm PL}$        & $0.04_{0.01}^{0.02}$  & --     \\
$F_{\mathrm{PL}}(\times 10^{-9})$  & $0.273_{-0.006}^{+0.005}$ & -- \\
\hline
$\chi^2 / {\rm d.o.f.}$ & $20.56/22$      & 12.86/11     \\
\hline
\end{tabular}
\label{tab:pers_cutoff}
\end{center}
\end{table}

\clearpage

\begin{table}
\begin{center}
\caption{Results from fitting the persistent X-ray spectrum to a more physical (\textsc{nthcomp+powerlaw}) model. For both the \textsc{nthcomp} (nthcomp) and \textsc{powerlaw} (PL) model, the flux, $F$, is in units of erg\,cm$^{-2}$\,s$^{-1}$, $\Gamma_{\rm X}$ is the photon index of the powerlaw, and $K$ is the normalisation in photons\,keV$^{-1}$\,cm$^{-2}$\,s$^{-1}$ at 1\,keV. For \textsc{nthcomp} alone, $kT_{\rm e}$ is the electron temperature in keV, while $kT_{\rm bb}$ is the seed photon temperature in keV.}
\begin{tabular}{ l c c }
Model: & \multicolumn{2}{c}{\textsc{nthcomp+powerlaw}} \\
\hline
Parameter      & 4U1728 & 4U1636 \\
\hline
$\Gamma_{\rm X,nthcomp}$      & $2.2_{-0.2}^{+0.1}$ & $1.90_{-0.06}^{+0.06}$ \\
$kT_{\rm e}$           & $6_{-1}^{+2}$    & $31_{-6}^{+13}$   \\
$kT_{\rm bb}$           & ${0.1}$    & ${0.1}$   \\
$K_{\rm nthcomp}$        & $2.0_{-0.2}^{+0.2}$  & $0.31_{-0.03}^{+0.04}$    \\
$F_{\rm nthcomp}(\times 10^{-9})$         & $3.6_{-0.2}^{+0.2}$ & $2.0_{-0.1}^{+0.1}$\\
\hline
$\Gamma_{\rm X,PL}$       & $2.2_{-0.8}^{+0.3}$ & --\\
$K_{\rm PL}$        & $0.2_{-0.2}^{+0.5}$  & --     \\
$F_{\rm PL}(\times 10^{-9})$        & $0.5_{-0.5}^{+0.2}$ & --\\
\hline
$\chi^2 / {\rm d.o.f.}$ & $22.34/22$      & 8.94/11     \\
\hline

\end{tabular}
\label{tab:pers_nthc}
\end{center}
\end{table}

\clearpage
\newpage

\begin{center}
\footnotesize
\begin{longtable}{ccccccccc}
\caption{Time resolved spectroscopy of the 4U1728 X-ray bursts. For each burst spectra were extracted in three time bins. Burst 9 was excluded from fitting due to artefacts in the X-ray data immediately before and after the burst.} \label{tab:resolved_1728} \\

\hline \multicolumn{1}{c}{Burst} & \multicolumn{1}{c}{Start} & \multicolumn{1}{c}{Stop}  & \multicolumn{1}{c}{Exp.}	&\multicolumn{1}{c}{$kT_{\rm bb}$}	&\multicolumn{1}{c}{$R_{\rm km}^2/D_{10}^2$}  & \multicolumn{1}{c}{Flux}			 & \multicolumn{1}{c}{$\chi^2$/d.o.f.}  \\

\multicolumn{1}{c}{} & \multicolumn{1}{c}{(MJD)}		  &\multicolumn{1}{c}{(MJD)}		&\multicolumn{1}{c}{(sec)}	  &\multicolumn{1}{c}{(keV)}	  	   & \multicolumn{1}{c}{}             &\multicolumn{1}{c}{$\times10^{-8}$}    &\\
\multicolumn{1}{c}{}& \multicolumn{1}{c}{}		  &	\multicolumn{1}{c}{}	&	\multicolumn{1}{c}{}  &	\multicolumn{1}{c}{}  	   & \multicolumn{1}{c}{}             &\multicolumn{1}{c}{(erg/cm$^2$/s)}    & \multicolumn{1}{c}{} \\

\hline

\endfirsthead

\multicolumn{8}{c}%
{{\bfseries \tablename\ \thetable{} -- continued from previous page}} \\

\hline \multicolumn{1}{c}{Burst} & \multicolumn{1}{c}{Start} & \multicolumn{1}{c}{Stop}  & \multicolumn{1}{c}{Exp.}	&\multicolumn{1}{c}{$kT_{\rm bb}$}	&\multicolumn{1}{c}{$R_{\rm km}^2/D_{10}^2$}  & \multicolumn{1}{c}{Flux}			 & \multicolumn{1}{c}{$\chi^2$/d.o.f.}  \\

\multicolumn{1}{c}{} & \multicolumn{1}{c}{(MJD)}		  &\multicolumn{1}{c}{(MJD)}		&\multicolumn{1}{c}{(sec)}	  &\multicolumn{1}{c}{(keV)}	  	   & \multicolumn{1}{c}{}             &\multicolumn{1}{c}{$\times10^{-8}$}    &\\
\multicolumn{1}{c}{}& \multicolumn{1}{c}{}		  &	\multicolumn{1}{c}{}	&	\multicolumn{1}{c}{}  &	\multicolumn{1}{c}{}  	   & \multicolumn{1}{c}{}             &\multicolumn{1}{c}{(erg/cm$^2$/s)}    & \multicolumn{1}{c}{} \\

\hline 
\endhead

\hline \multicolumn{8}{r}{{Continued on next page}} \\ \hline
\endfoot

\hline 
\endlastfoot

B1	    &59,307.719803    &59,307.719837   &2.38  &$2.3_{-0.7}^{+0.9}$    &$195_{-68}^{+68}$     &$6_{-2}^{+2}$   &1.76/4\\
	&59,307.719837    &59,307.719895   &3.95  &$1.5_{-0.5}^{+0.6}$    &$606_{-249}^{+249}$   &$4_{-2}^{+1}$   &1.89/4\\
	&59,307.719895    &59,307.720011   &7.97  &--			    &--			   &--		    &--\\
 \hline	                              
B2	    &59,307.899837    &59,307.899872   &2.33  &$2.7_{-0.6}^{+0.8}$    &$134_{-35}^{+35}$     &$8_{-2}^{+2}$   &0.54/4\\
	&59,307.899872    &59,307.899930   &3.89  &$2.4_{-0.6}^{+0.7}$    &$113_{-32}^{+32}$     &$4_{-1}^{+1}$   &0.40/4\\
	&59,307.899930    &59,307.900046   &7.98  &$1.5_{-0.6}^{+0.7}$    &$160_{-71}^{+71}$     &$0.9_{-0.4}^{+0.4}$ &1.30/4\\
 \hline	                              
B3	    &59,308.035578    &59,308.035613   &2.35  &$2.2_{-0.6}^{+0.7}$    &$288_{-80}^{+80}$     &$8_{-2}^{+2}$   &1.04/4\\
	&59,308.035613    &59,308.035671   &3.91  &$2.5_{-0.7}^{+0.7}$    &$104_{-35}^{+35}$     &$4_{-1}^{+1}$   &2.35/4\\
	&59,308.035671    &59,308.035787   &8.00  & $<2$   &$<11$         &$0.4_{-1 }^{+1}$    &1.14/2\\
 \hline	                              
B4	    &59,308.163425    &59,308.163460   &2.38  &$2.6_{-0.7}^{+0.8}$    &$132_{-45}^{+45}$     &$7_{-2}^{+2}$   &0.93/4\\
	&59,308.163460    &59,308.163518   &3.94  &$2.4_{-0.8}^{+1.4}$    &$86_{-38}^{+38}$      &$3_{-1}^{+1}$   &0.89/4\\
	&59,308.163518    &59,308.163634   &7.99  &$<4.2$    &$137_{-97}^{+97}$     &$0.7_{-0.5}^{+1}$   &1.23/3\\
 \hline	                              
B5	    &59,308.281990    &59,308.282025   &2.37  &$2.6_{-0.9}^{+1.2}$    &$104_{-45}^{+45}$     &$5_{-2}^{+2}$   &1.13/4\\
	&59,308.282025    &59,308.282083   &3.97  &$2.7_{-1.0}^{+1.7}$    &$62_{-27}^{+27}$      &$4_{-2}^{+2}$   &0.62/4\\
	&59,308.282083    &59,308.282199   &8.00  &$<3.3$     &--                    &--                 &--\\
 \hline	                              
B6	    &59,308.401388    &59,308.401423   &2.38  &$2.5_{-0.7}^{+1.1}$    &$93_{-35}^{+35}$      &$4_{-1}^{+1}$   &0.27/4\\
	&59,308.401423    &59,308.401481   &3.93  &$2.2_{-0.5}^{+0.7}$    &$152_{-44}^{+44}$     &$4_{-1}^{+1}$   &0.09/4\\
	&59,308.401481    &59,308.401597   &7.98  &$<4.1$    &$100_{-75}^{+75}$     &$0.5_{-0.3}^{+1.0}$   &0.85/3\\
 \hline	                              
B7	    &59,308.563761    &59,308.563796   &2.37  &$3.0_{-0.8}^{+1.4}$    &$75_{-26}^{+26}$      &$7_{-2}^{+3}$   &1.27/4\\
	&59,308.563796    &59,308.563854   &3.97  &$1.9_{-0.6}^{+0.9}$    &$246_{-94}^{+94}$     &$4_{-1}^{+1}$   &2.35/4\\
	&59,308.563854    &59,308.563969   &7.95  &--                     &--                    &--                 &--\\
\hline	                              
B8	    &59,308.732175    &59,308.732210   &2.35  &$2.0_{-0.5}^{+0.5}$    &$368_{-100}^{+100}$   &$7_{-2}^{+2}$   &1.25/4\\
	&59,308.732210    &59,308.732268   &3.94  &$1.7_{-0.6}^{+0.6}$    &$450_{-127}^{+127}$   &$4_{-1}^{+1}$   &2.94/4\\
	&59,308.732268    &59,308.732384   &7.96  &$1.3_{-1.3}^{+1.5}$    &$207_{-135}^{+135}$   &$0.8_{-0.4}^{+1}$  &0.54/3\\
 \hline	                              
B9	    &59,308.887569    &59,308.887604   &2.37  &$2.2_{-0.7}^{+0.9}$    &$224_{-82}^{+82}$     &$6_{-2}^{+2}$   &0.40/4\\
	&59,308.887604    &59,308.887662   &3.95  &$2.5_{-0.6}^{+0.8}$    &$112_{-34}^{+34}$     &$5_{-2}^{+2}$   &0.79/4\\
	&59,308.887662    &59,308.887777   &7.95  &--		     &--		    &--			&--\\
 \hline	                              
B10	    &59,309.029884    &59,309.029918   &2.36  &$2.3_{-0.6}^{+0.8}$    &$209_{-60}^{+60}$     &$6_{-2}^{+2}$   &1.40/4\\
	&59,309.029918    &59,309.029976   &3.96  &$2.6_{-0.8}^{+1.1}$    &$49_{-21}^{+21}$      &$2_{-1}^{+1}$   &0.56/4\\
	&59,309.029976    &59,309.030092   &7.97  &$1.0_{-1.0}^{+0.8}$    &$844_{-512}^{+512}$   &$<2$	&0.45/3\\
 \hline	                              
B11	    &59,309.277905    &59,309.277939   &2.38  &$<4.8$    &$91_{-47}^{+47}$      &$3_{-2}^{+3}$   &2.04/4\\
	&59,309.277939    &59,309.277997   &3.97  &$1.8_{-0.7}^{+0.8}$    &$172_{-80}^{+80}$     &$2_{-1}^{+1}$   &1.14/4\\
	&59,309.277997    &59,309.278113   &7.98  &$<2.6$   &$<299$   &$<2$  &0.62/3\\
 \hline	                              
B12	    &59,309.426932    &59,309.426967   &2.33  &$2.7_{-0.6}^{+0.8}$    &$134_{-35}^{+35}$     &$8_{-2}^{+2}$   &0.54/4\\
	&59,309.426967    &59,309.427025   &3.93  &$2.0_{-0.5}^{+0.6}$    &$228_{-71}^{+71}$     &$4_{-1}^{+1}$   &1.07/4\\
	&59,309.427025    &59,309.427141   &7.92  &$<3.6$    &$45_{-27}^{+27}$      &$<2$   &2.39/3\\
 \hline	                              
B13	    &59,309.596493    &59,309.596527   &2.37  &$2.5_{-0.9}^{+1.4}$    &$89_{-38}^{+38}$      &$4_{-2}^{+2}$   &0.90/4\\
	&59,309.596527    &59,309.596585   &3.93  &$2.1_{-0.5}^{+0.6}$    &$287_{-78}^{+78}$     &$5_{-1}^{+2}$   &1.85/4\\
	&59,309.596585    &59,309.596701   &7.95  &$1.5_{-0.7}^{+1.0}$    &$211_{-108}^{+108}$   &$<2$   &0.22/3\\
 \hline	                              
B14	    &59,309.729525    &59,309.729560   &2.38  &$2.2_{-0.9}^{+1.1}$    &$183_{-74}^{+74}$     &$4_{-2}^{+2}$   &0.85/4\\
	&59,309.729560    &59,309.729618   &3.98  &$1.4_{-0.6}^{+0.9}$    &$495_{-218}^{+218}$   &$2_{-1}^{+2}$   &1.51/4\\
	&59,309.729618    &59,309.729733   &7.98  &--		     &--		    &--			&--\\					      
\end{longtable}
\end{center}

\newpage

\begin{table}
\begin{center}
\caption{Time-lag measurements for the stacked signals of 4U1728. Here ${\xi_2-\xi_1=\left(\frac{1}{\nu_2}\right)-\left(\frac{1}{\nu_1}\right)}$.}
\begin{tabular}{ c c c c }
\hline
$\nu_1$&  $\nu_2$   & $\xi_2-\xi_1$ & Lag (min) \\[0.05cm]
\hline
$1.2\times10^{18}$\,GHz/5 keV&   9\,GHz & 0.11 & $2.5^{+0.7}_{-1.1}$ \\[0.05cm]
$1.2\times10^{18}$\,GHz/5 keV&   5.5\,GHz & 0.18 & $3.5^{+1.2}_{-0.7}$ \\[0.05cm]
9\,GHz&   5.5\,GHz & 0.07 & $1.0^{+1.0}_{-0.9}$ \\
\hline

\hline
\end{tabular}
\label{tab:stacked_lags}
\end{center}
\end{table}
\clearpage

\begin{figure}
   \centering
      { \includegraphics[width=1\columnwidth]{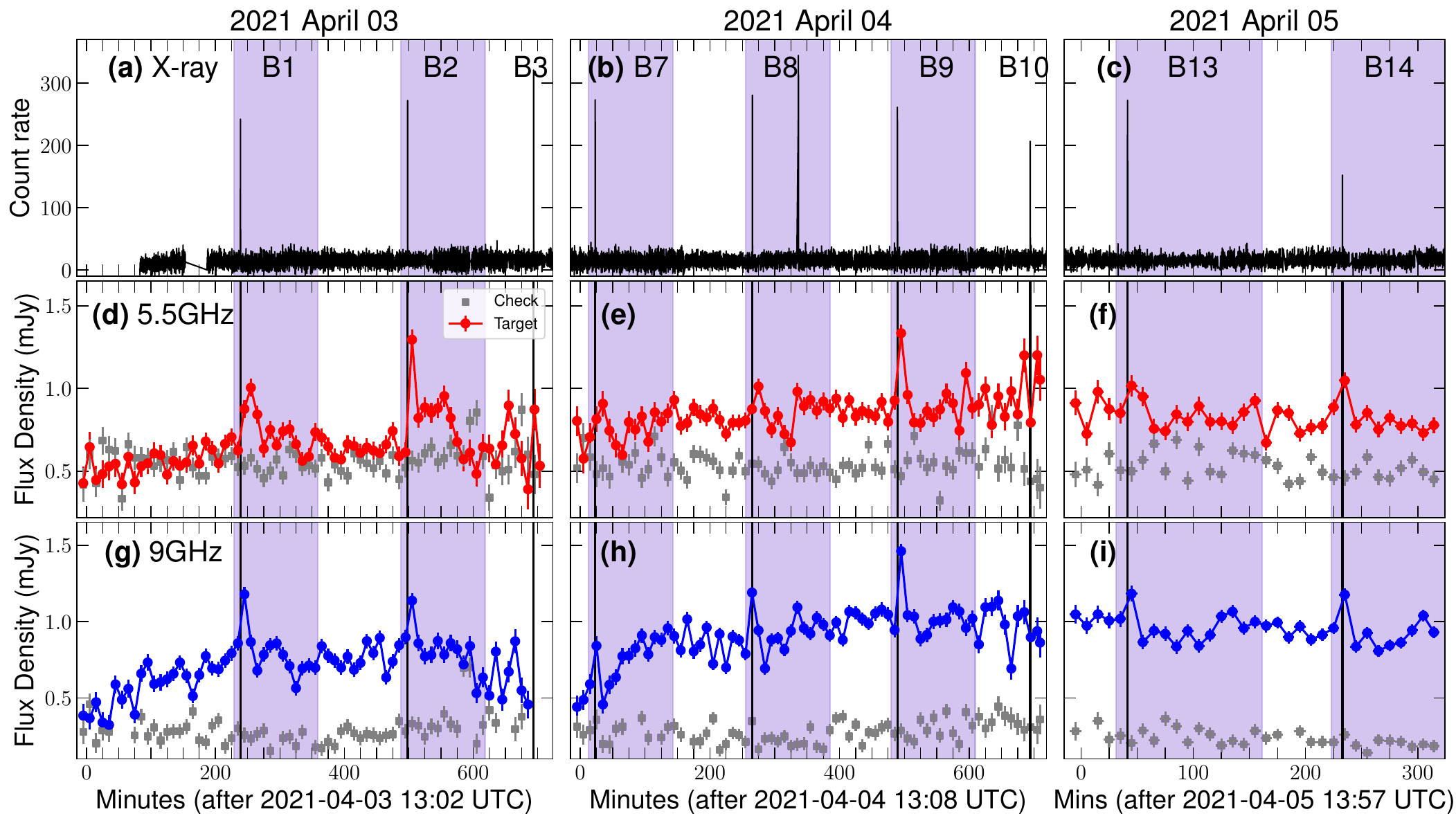}}\\

 \caption{\label{fig:bursttimes} X-ray and multi-band radio light curves of 4U1728 on 10-minute timescales for our three radio epochs. For the X-rays we show the 1-second 3--25\,keV light curves (panels a, b, s). For the radio, we show the flux densities of the target at 5.5\,GHz (panels d, e, f; red circles) and 9\,GHz (panels g, h, i; blue circles), as well as a nearby check source (grey squares), where error bars are 1-sigma. We detected 10 X-ray bursts coincident with our radio monitoring. For all X-ray bursts, we find clearly defined radio counterparts, with the only exceptions being two that occurred right at the end of our observation when the source elevation was low and the observation ended during the 10-minute time bin, however, radio brightening is detected on shorter timescales for these two bursts (Fig.~\ref{fig:bursts_4min}). No radio brightening was detected in the nearby check source. Burst times are shown by the vertical black lines in the lower two rows, while the vertical shaded purple regions mark the time chunks used for our cross-correlation analysis (see Section~\ref{sec:timelags}) of each burst, where the bursts are labelled as in Extended Data Table~\ref{tab:resolved_1728} (we exclude the two bursts close to the end of epoch 1 and~2, labelled as B3 and B11, from our cross-correlation analysis of single bursts (but not for the stacked analysis), as such there is no vertical shaded region for these two bursts).
 }
 \end{figure}

\begin{figure}
   \centering
   
   { \includegraphics[width=0.6\columnwidth]{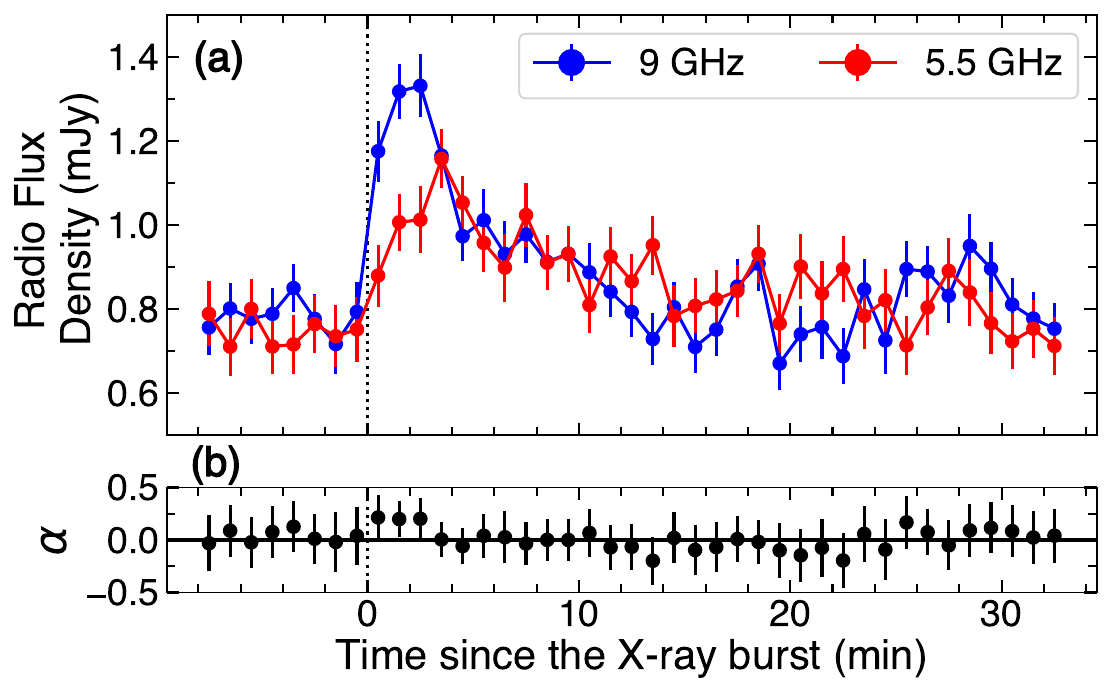}}\\

 \caption{\label{fig:radiostack} Stacked radio light curves and time-resolved radio spectral indices of 4U1728. To create these light curves (panel a), we stacked all of the radio bursts from 4U1728 separately in the 5.5\,GHz and 9.0\,GHz bands, with respect to the time of the partner X-ray burst (indicated by the vertical black dotted line). Panel b displays the stacked spectral index showing the flare starting as optically thick. There is a clear time-lag between the X-ray and radio burst signals, where the radio signal lags the X-ray signal by several minutes and this lag is dependent on electromagnetic frequency (smaller electromagnetic frequencies yield longer lags). All errors are 1-sigma.
 }
 \end{figure}

\clearpage

 \begin{figure}
   \centering
   
   { \includegraphics[width=1\columnwidth]{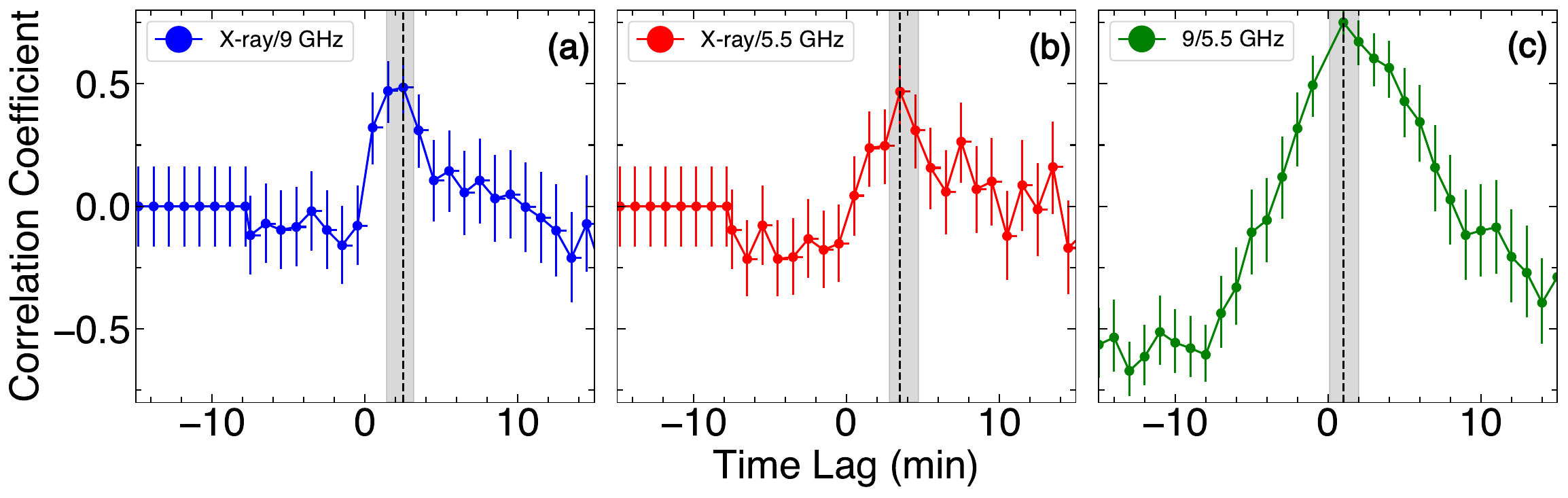}}
   \\

 \caption{\label{fig:lag_zdcf} Cross-correlation functions created with the ZDCF algorithm from stacked versions of the X-ray and radio signals; X-ray to 9\,GHz radio (panel a), X-ray to 5.5\,GHz radio (panel b), and 9\,GHz to 5.5\,GHz radio (panel c), with 1-sigma errors. The resulting lag measurements and their uncertainties are shown with the dotted black lines and shaded gray regions, respectively. The lag measurements from these CCFs are also tabulated in Extended Data Table~\ref{tab:stacked_lags}.
 }
 \end{figure}
\clearpage

\begin{figure}
   \centering
   
   { \includegraphics[width=1\columnwidth]{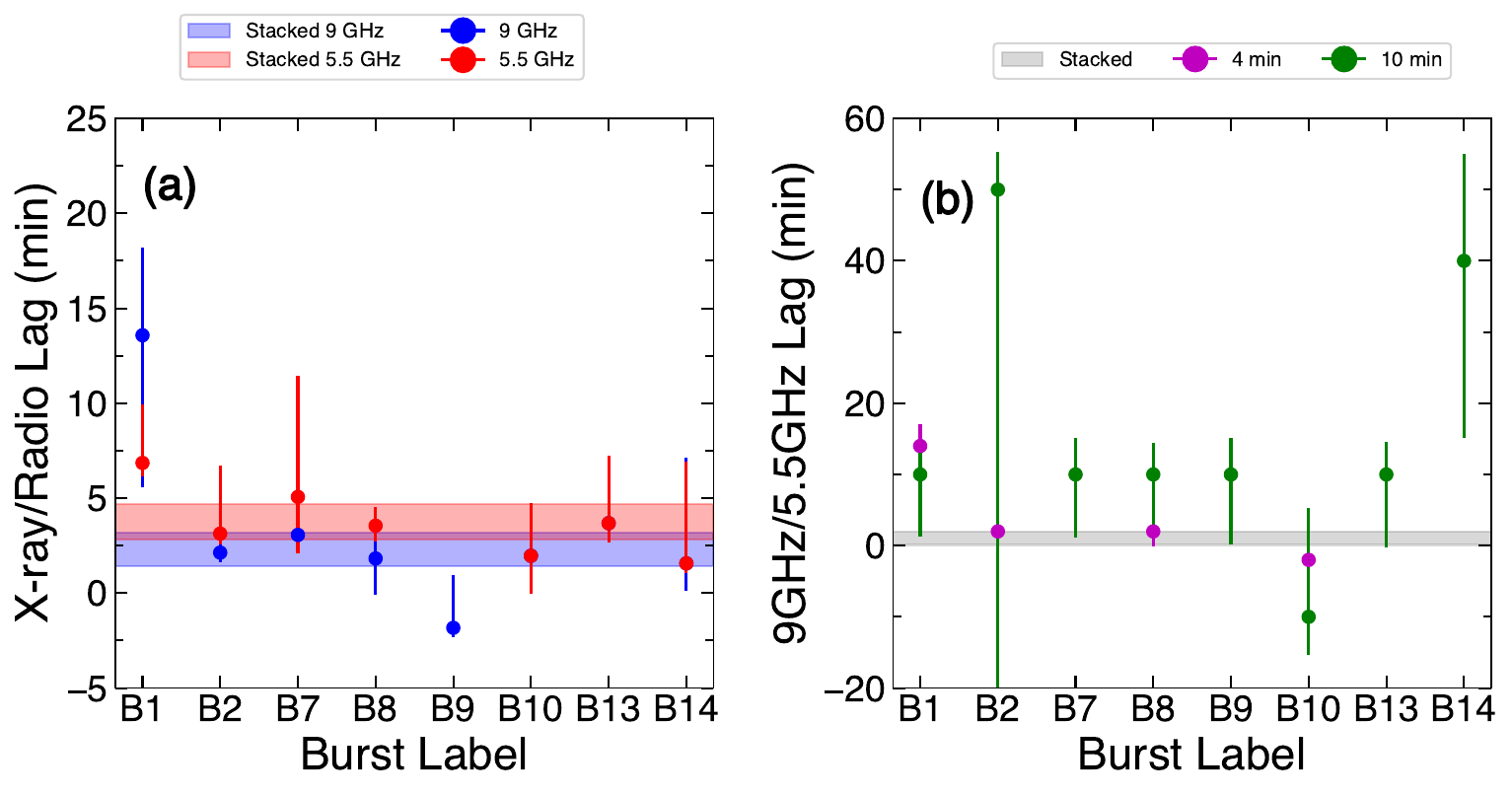}}\\

 \caption{\label{fig:lags} Measured time-lags between the X-ray and radio bands in 4U1728. Here we show the lags between the X-ray and stacked signal, as well as those for individual bursts (panel a), as well as the lag between the 9 and 5.5\,GHz signal (panel b). In both panels, the data points represent lags measured from cross-correlation functions comparing signals from individual bursts, while the shaded regions represent lags measured from cross-correlation functions comparing signals from the stacked bursts. We show the results of both 2\,min and 10\,min time-binned radio light curves for the 9\,GHz to 5.5\,GHz radio lags (which are consistent within errors with each other), while we show only the results from the 10\,min time-binned radio light curves for the X-ray to radio lags for clarity. In both cases, the individual burst lags are consistent with the stacked measurement within (the 1-sigma) uncertainties.
 }
 \end{figure}

\clearpage

\begin{figure}
   \centering
   
   { \includegraphics[width=0.55\columnwidth]{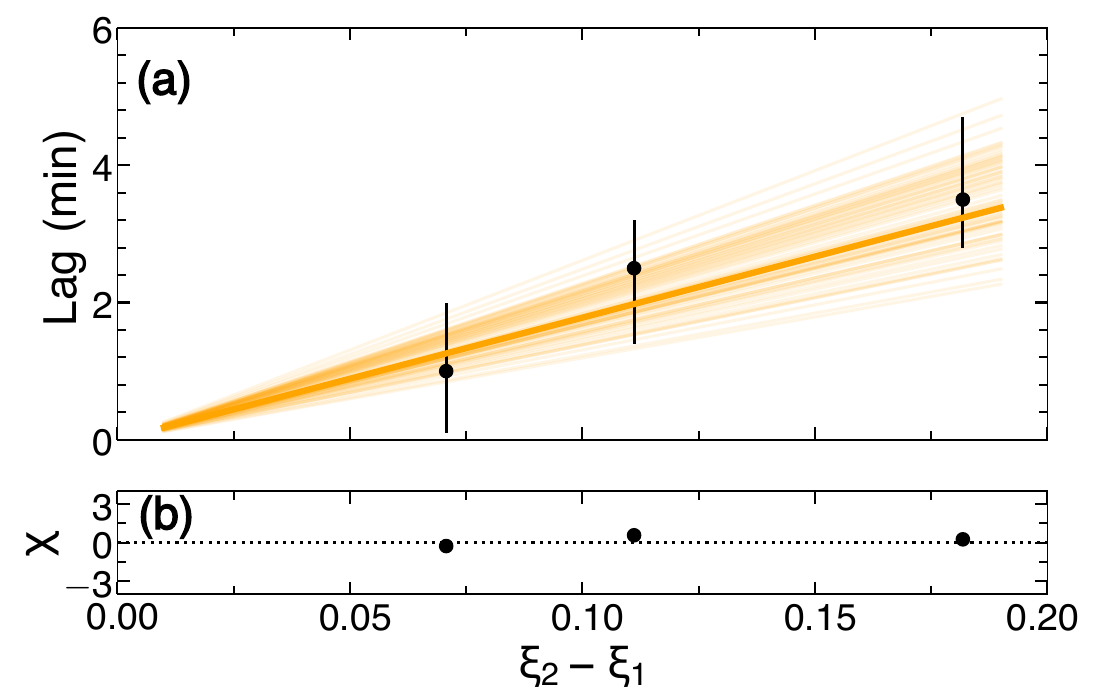}}
   { \includegraphics[width=0.42\columnwidth]{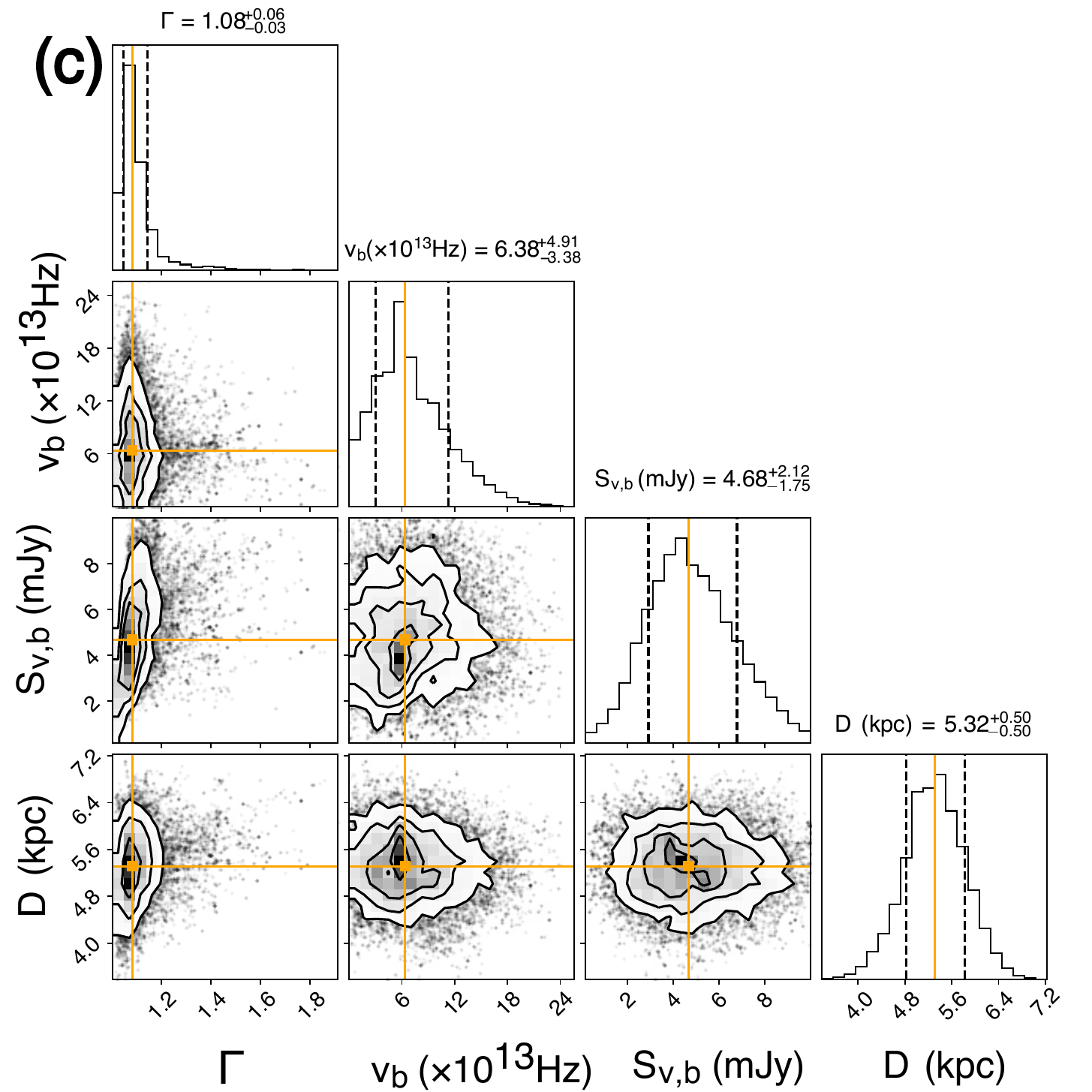}}\\

 \caption{\label{fig:lagfit} Results of modelling the stacked burst time-lags from 4U1728. Panel a: The best-fit model (solid orange line) overlaid on the time-lag measurements, with 1-sigma errors. Here the shaded orange region represents the model predictions from the final position of the MCMC walkers in our parameter space. The three lag measurements correspond to lags between the X-ray and the 9\,GHz and 5\,GHz radio data, as well as the lag between the 9\,GHz and 5.5\,GHz radio data (as shown in Extended Data Figure~\ref{fig:lag_zdcf} and Extended Data Table~\ref{tab:stacked_lags}). Panel b: The residuals of the fit (data--model/uncertainties). Panel c: Two-parameter correlations in our modelling for the free parameters and those with informative (non-uniform) priors. Our model fits the measured time-lags reasonably well, suggesting the radio counterparts to the X-ray bursts in 4U1728 originate in an out-flowing compact jet.
 }
 \end{figure}

\clearpage

\renewcommand{\tablename}{Supplementary Table}
\setcounter{table}{0}

\begin{center}
\footnotesize
\begin{longtable}{ccccccccc}
\caption{Time resolved spectroscopy of the 4U1636 X-ray bursts. For each burst, spectra were extracted in five time bins.} \label{tab:resolved_1636} \\

\hline \multicolumn{1}{c}{Burst} & \multicolumn{1}{c}{Start} & \multicolumn{1}{c}{Stop}  & \multicolumn{1}{c}{Exp.}	&\multicolumn{1}{c}{$kT_{\rm bb}$}	&\multicolumn{1}{c}{$R_{\rm km}^2/D_{10}^2$}  & \multicolumn{1}{c}{Flux}			 & \multicolumn{1}{c}{$\chi^2$/d.o.f.}  \\

\multicolumn{1}{c}{} & \multicolumn{1}{c}{(MJD)}		  &\multicolumn{1}{c}{(MJD)}		&\multicolumn{1}{c}{(sec)}	  &\multicolumn{1}{c}{(keV)}	  	   & \multicolumn{1}{c}{}             &\multicolumn{1}{c}{$\times10^{-8}$}    &\\
\multicolumn{1}{c}{}& \multicolumn{1}{c}{}		  &	\multicolumn{1}{c}{}	&	\multicolumn{1}{c}{}  &	\multicolumn{1}{c}{}  	   & \multicolumn{1}{c}{}             &\multicolumn{1}{c}{(erg/cm$^2$/s)}    & \multicolumn{1}{c}{} \\

\hline

\endfirsthead

\multicolumn{8}{c}%
{{\bfseries \tablename\ \thetable{} -- continued from previous page}} \\

\hline \multicolumn{1}{c}{Burst} & \multicolumn{1}{c}{Start} & \multicolumn{1}{c}{Stop}  & \multicolumn{1}{c}{Exp.}	&\multicolumn{1}{c}{$kT_{\rm bb}$}	&\multicolumn{1}{c}{$R_{\rm km}^2/D_{10}^2$}  & \multicolumn{1}{c}{Flux}			 & \multicolumn{1}{c}{$\chi^2$/d.o.f.}  \\

\multicolumn{1}{c}{} & \multicolumn{1}{c}{(MJD)}		  &\multicolumn{1}{c}{(MJD)}		&\multicolumn{1}{c}{(sec)}	  &\multicolumn{1}{c}{(keV)}	  	   & \multicolumn{1}{c}{}             &\multicolumn{1}{c}{$\times10^{-8}$}    &\\
\multicolumn{1}{c}{}& \multicolumn{1}{c}{}		  &	\multicolumn{1}{c}{}	&	\multicolumn{1}{c}{}  &	\multicolumn{1}{c}{}  	   & \multicolumn{1}{c}{}             &\multicolumn{1}{c}{(erg/cm$^2$/s)}    & \multicolumn{1}{c}{} \\

\hline 
\endhead

\hline \multicolumn{8}{r}{{Continued on next page}} \\ \hline
\endfoot

\hline
\endlastfoot
B1  &59,491.256215   &59,491.256261    &3.20   &$1.9_{-0.6}^{+0.7}$   &$165_{-71}^{+71}$	 &$2_{-1}^{+1}$    &1.28/3 \\
   &59,491.256261   &59,491.256307    &3.19   &$1.9_{-0.8}^{+1.0}$    &$167_{-86}^{+86}$   	 &$2_{-2}^{+2}$    &1.98/3 \\
   &59,491.256307   &59,491.256400    &6.39   &$1.8_{-0.7}^{+0.8}$    &$171_{-68}^{+68}$   	 &$1.7_{-0.7}^{+0.7}$  	  &0.31/3 \\
   &59,491.256400   &59,491.256585    &12.88  &$1.9_{-1.0}^{+1.7}$    &$34_{-20}^{+20}$   	 &$0.5_{-0.2}^{+1.3}$ 	  &0.42/3 \\
   &59,491.256585   &59,491.256956    &25.77  &$1.8_{-0.8}^{+1.3}$    &$34_{-18}^{+18}$       &$0.3_{-0.2}^{+0.2}$     &1.26/3 \\
   &59,491.256956   &59,491.257696    &51.65  &$<2.4$    &--		         &--		    &--		    \\
\hline
B2   &59,491.450034   &59,491.450081    &3.19   &$2.1_{-0.8}^{+1.3}$    &$116_{-50}^{+49}$   	 &$2_{-1}^{+1}$        &0.85/3 \\ 
   &59,491.450081   &59,491.450127    &3.19   &$1.9_{-0.7}^{+1.0}$    &$130_{-68}^{+68}$   	  &$1.5_{-0.7}^{+8.9}$	&2.12/3 \\ 
   &59,491.450127   &59,491.450219    &6.40   &$1.7_{-0.6}^{+0.8}$    &$140_{-62}^{+62}$   	  &$1.2_{-0.6}^{+0.6}$ 	&0.56/3 \\
   &59,491.450219   &59,491.450405    &12.83  &$1.7_{-0.6}^{+0.7}$    &$119_{-46}^{+46}$   	  &$0.9_{-0.4}^{+0.3}$ 	&0.10/3 \\
   &59,491.450405   &59,491.450775    &25.73  &$1.7_{-0.7}^{+1.1}$    &$35_{-21}^{+20}$        &$0.3_{-0.2}^{+0.2}$  &1.03/3 \\
   &59,491.450775   &59,491.451516    &51.55  &$<2.6$    &$<81$        &$<0.2$	&1.02/3   \\
\hline
B3   &59,491.791631 &59,491.791678    &3.19   &$1.4_{-0.7}^{+1.1}$    &$433_{-208}^{+208}$   	 &$2_{-9}^{+1}$        &0.69/3 \\
   &59,491.791678   &59,491.791724    &3.19   &$1.3_{-0.9}^{+1.3}$    &$851_{-399}^{+400}$     &$2.1_{0.9}^{+1.3}$	&2.23/3 \\
   &59,491.791724   &59,491.791817    &6.40   &$1.6_{-0.5}^{+0.7}$    &$232_{-89}^{+88}$   	  &$1.7_{-0.7}^{+0.6}$ 	&0.54/3 \\
   &59,491.791817   &59,491.792002    &12.86  &$1.7_{-0.6}^{+0.7}$    &$101_{-42}^{+42}$   	  &$0.9_{-0.4}^{+0.3}$  &0.55/3 \\
   &59,491.792002   &59,491.792372    &25.73  &$1.6_{-0.7}^{+0.9}$    &$56_{-27}^{+26}$        &$0.4_{-0.2}^{+0.2}$  &0.80/3 \\
   &59,491.792372   &59,491.793113    &51.55  &$<6.4$    &$<2$           &$<0.2$ 	&1.12/3   \\
\hline
B4   &59,492.095034   &59,492.095081    &3.21   &$<3$    &$339_{-162}^{+162}$   	 &$2_{-1}^{+1}$    &1.44/3 \\
   &59,492.095081   &59,492.095127    &3.21   &$2.8_{-1.2}^{+3.6}$    &$36_{-18}^{+18}$   	 &$2_{-1}^{+2}$ 	&0.13/3 \\
   &59,492.095127   &59,492.095219    &6.42   &$2.1_{-0.9}^{+1.7}$    &$53_{-28}^{+28}$   	 &$1.2_{-0.7}^{+0.7}$ 	  &0.11/3 \\
   &59,492.095219   &59,492.095405    &12.87  &$1.6_{-0.7}^{+0.9}$    &$115_{-53}^{+53}$   	 &$0.7_{-0.3}^{+0.3}$	  &0.34/3 \\
   &59,492.095405   &59,492.095775    &25.78  &$1.8_{-0.9}^{+1.7}$    &$29_{-16}^{+15}$            &$0.3_{-0.2}^{+0.2}$     &1.42/3 \\
   &59,492.095775   &59,492.096516    &51.65  &$<4.1$    &$<11$               &$<0.2$ 	&0.57/3   \\
\hline
B5   &59,492.246238   &59,492.246284    &3.18   &$1.9_{-0.5}^{+0.7}$    &$184_{-66}^{+65}$   	 &$3_{-1}^{+1}$    &0.28/3 \\
   &59,492.246273   &59,492.246319    &3.17   &$2.2_{-0.7}^{+0.9}$    &$109_{-41}^{+40}$   	 &$5_{-1}^{+1}$ 	&0.53/3 \\ 
   &59,492.246319   &59,492.246412    &6.36   &$1.8_{-0.5}^{+0.6}$    &$170_{-58}^{+58}$   	 &$1.7_{-0.5}^{+0.7}$  	  &0.37/3 \\
   &59,492.246412   &59,492.246597    &12.80  &$1.9_{-0.9}^{+1.7}$    &$49_{-22}^{+22}$   	 &$0.8_{-0.3}^{+1.3}$  	  &1.36/3 \\
   &59,492.246597   &59,492.246967    &25.74  &$1.5_{-0.7}^{+0.9}$    &$63_{-31}^{+30}$            &$0.3_{-0.1}^{+0.1}$     &1.27/3 \\
\hline
B6   &59,492.399918   &59,492.399965    &3.20   &$2.3_{-0.8}^{+1.2}$    &$103_{-41}^{+41}$   	 &$3_{-1}^{+1}$    &0.56/3 \\
   &59,492.399965   &59,492.400011    &3.20   &$1.2_{-0.6}^{+1.2}$    &$875_{-403}^{+404}$   	 &$2.9_{-0.9}^{+1.9}$ 	  &1.83/3 \\
   &59,492.400011   &59,492.400104    &6.39   &$2.3_{-0.9}^{+1.7}$    &$40_{-20}^{+20}$   	 &$1.2_{-0.6}^{+1.7}$ 	  &0.58/3 \\
   &59,492.400104   &59,492.400289    &12.83  &$1.6_{-0.9}^{+1.1}$    &$85_{-40}^{+40}$   	 &$0.6_{-0.3}^{+1.3}$  	  &0.44/3 \\
   &59,492.400289   &59,492.400659    &25.75  &$1.6_{-0.7}^{+1.0}$    &$44_{-23}^{+23}$            &$0.3_{-0.2}^{+0.2}$     &1.01/3 \\
   &59,492.400659   &59,492.401400    &51.57  &$<3$    &$448_{-276}^{+277}$         &$0.2_{-0.1}^{+1.0}$ 	&1.41/3   \\
\hline
B7   &59,492.550995   &59,492.551041    &3.18   &$2.9_{-1.0}^{+2.0}$    &$34_{-14}^{+14}$   	 &$3_{-1}^{+1}$    &0.38/3 \\
   &59,492.551041   &59,492.551087    &3.18   &$2.0_{-0.7}^{+0.9}$    &$115_{-45}^{+45}$   	 &$2.1_{-0.9}^{+0.9}$ 	  &0.36/3 \\
   &59,492.551087   &59,492.551180    &6.39   &$1.9_{-0.7}^{+1.0}$    &$80_{-37}^{+37}$   	 &$1.0_{-0.5}^{+1.5}$ 	  &0.64/3 \\
   &59,492.551180   &59,492.551365    &12.79  &$1.7_{-0.6}^{+0.7}$    &$75_{-27}^{+27}$   	 &$0.8_{-0.3}^{+0.2}$ 	  &0.49/3 \\
   &59,492.551365   &59,492.551736    &25.70  &$1.3_{-0.9}^{+2.3}$    &$68_{-46}^{+45}$        	 &$0.2_{-0.1}^{+0.5}$     &1.07/3 \\
   &59,492.551736   &59,492.552476    &51.38  &$<0.8$    &--			 &--		       &--	    \\
\hline
B8   &59,492.700995   &59,492.701041    &3.19   &$1.9_{-0.6}^{+0.9}$    &$131_{-56}^{+56}$   	 &$2_{-1}^{+8}$    &2.49/3 \\
   &59,492.701041   &59,492.701087    &3.19   &$1.8_{-0.6}^{+0.7}$    &$196_{-80}^{+79}$   	 &$2.1_{-0.9}^{+1.0}$ 	  &0.96/3 \\
`   &59,492.701087   &59,492.701180    &6.39   &$2.2_{-0.7}^{+1.0}$    &$70_{-23}^{+23}$   	 &$1.9_{-0.6}^{+1.7}$	  &0.43/3 \\
   &59,492.701180   &59,492.701365    &12.83  &$1.2_{-0.5}^{+0.6}$    &$420_{-171}^{+171}$   	 &$0.8_{-0.4}^{+0.5}$  	  &1.38/3 \\
   &59,492.701365   &59,492.701736    &25.70  &$1.0_{-1.0}^{+1.0}$    &$292_{-146}^{+146}$    	 &$0.3_{-0.2}^{+0.8}$     &2.33/3 \\
   &59,492.701736   &59,492.702476    &51.50  &$<3.5$   &$40_{-28}^{+27}$        	 &$0.8_{-0.1}^{+0.1}$     &0.98/3   \\
\hline
B9   &59,492.887604   &59,492.887650    &3.18   &$1.6_{-0.5}^{+0.6}$    &$330_{-117}^{+117}$  	 &$2_{-1}^{+7}$    &0.66/3 \\
   &59,492.887650   &59,492.887696    &3.17   &$1.9_{-0.5}^{+0.6}$    &$207_{-61}^{+61}$  	 &$3.0_{-1}^{+0.9}$ 	  &0.51/3 \\
   &59,492.887696   &59,492.887789    &6.35   &$2.3_{-0.8}^{+1.2}$    &$52_{-18}^{+17}$   	 &$1.5_{-0.5}^{+1.7}$ 	  &1.15/3 \\
   &59,492.887789   &59,492.887974    &12.80  &$1.4_{-0.4}^{+0.7}$    &$182_{-62}^{+62}$       	 &$0.8_{-0.3}^{+0.2}$     &2.02/3 \\
   &59,492.887974   &59,492.888344    &25.69  &$1.4_{-0.6}^{+0.8}$    &$108_{-44}^{+44}$       	 &$0.4_{-0.2}^{+0.2}$     &0.42/3 \\
   &59,492.888344   &59,492.889085    &51.42  &$1.2_{-1.2}^{+1.1}$    &$58_{-36}^{+36}$        	 &$0.11_{-0.07}^{+0.11}$ &0.08/3   \\
\hline
B10   &59,493.035509   &59,493.035555    &3.18   &$1.8_{-0.6}^{+0.8}$    &$241_{-89}^{+89}$   	 &$3_{-1}^{+1}$    &0.51/3 \\
   &59,493.035555   &59,493.035601    &3.17   &$2.0_{-0.7}^{+1.3}$    &$133_{-53}^{+53}$   	 &$2.2_{-0.9}^{+0.9}$ 	  &2.72/3 \\
   &59,493.035601   &59,493.035694    &6.38   &$1.9_{-0.7}^{+0.8}$    &$96_{-38}^{+38}$   	 &$1.2_{-0.4}^{+0.5}$  	  &1.19/3 \\
   &59,493.035694   &59,493.035879    &12.84  &$1.6_{-0.7}^{+0.9}$    &$59_{-31}^{+31}$        	 &$0.4_{-0.2}^{+0.2}$     &1.44/3 \\
   &59,493.035879   &59,493.036250    &25.72  &$<2.9$    &$66_{-32}^{+32}$        	 &$0.3_{-0.1}^{+0.1}$     &1.81/3 \\
   &59,493.036250   &59,493.036990    &51.46  &$<2$    &$145_{-87}^{+87}$       	 &$0.21_{-0.09}^{+0.21}$ &0.69/3   \\
\hline
B11   &59,493.183449   &59,493.183495    &3.21   &$2.0_{-0.5}^{+0.7}$    &$164_{-64}^{+64}$   	 &$3_{-1}^{+1}$    &1.61/3 \\
   &59,493.183495   &59,493.183541    &3.18   &$2.6_{-2.6}^{+3.7}$    &$34_{-20}^{+19}$   	 &$3_{-1}^{+1}$  	  &1.76/3 \\
   &59,493.183541   &59,493.183634    &6.38   &$1.9_{-0.7}^{+1.0}$    &$77_{-37}^{+36}$  	 	 &$1.1_{-0.5}^{+1.5}$ 	  &0.43/3 \\
   &59,493.183634   &59,493.183819    &12.83  &$1.8_{-0.8}^{+1.3}$    &$62_{-27}^{+27}$        	 &$0.6_{-0.3}^{+0.3}$    &0.77/3 \\
   &59,493.183819   &59,493.184189    &25.71  &$<2.5$    &$98_{-48}^{+48}$        	 &$0.3_{-0.2}^{+0.2}$     &0.62/3 \\
   &59,493.184189   &59,493.184930    &51.51  &$<1.8$    &$1653_{-1306}^{+1311}$  	 &$0.22_{-0.10}^{+0.22}$ &0.29/3 \\
\hline
B12   &59,493.334872   &59,493.334918    &3.19   &$2.8_{-2.8}^{+9.3}$    &$22_{-15}^{+13}$   	 &$10_{-9}^{+7}$    &0.97/3 \\
   &59,493.334918   &59,493.334965    &3.19   &$2.0_{-0.9}^{+1.5}$    &$158_{-63}^{+62}$  	 &$4_{-2}^{+1}$ 	  &2.22/3 \\
   &59,493.334965   &59,493.335057    &6.37   &$1.8_{-0.6}^{+0.7}$    &$167_{-58}^{+58}$   	 &$2.0_{-0.6}^{+0.7}$ 	  &1.64/3 \\
   &59,493.335057   &59,493.335243    &12.83  &$1.7_{-0.7}^{+1.0}$    &$95_{-39}^{+39}$        	 &$0.7_{-0.3}^{+0.3}$     &0.68/3 \\
   &59,493.335243   &59,493.335613    &25.70  &$1.9_{-0.9}^{+1.8}$    &$22_{-11}^{+11}$        	 &$0.3_{-0.2}^{+0.2}$     &0.51/3 \\
   &59,493.335613   &59,493.336354    &51.46  &$<3.9$    &$46_{-41}^{+41}$         	 &$0.10_{-0.09}^{+0.1}$  &1.18/3 \\
\hline
\end{longtable}
\end{center}

\end{document}